\pdfoutput=1                     
\documentclass[
preprint,
amsmath,amssymb
]{revtex4-1}

\usepackage{graphicx}            

\usepackage[normalem]{ulem}
\usepackage[english]{babel}
\usepackage{color}
\usepackage{dcolumn}
\usepackage{bm}

\usepackage{booktabs}
\usepackage{float}

\usepackage[utf8]{inputenc}
\usepackage[T1]{fontenc}
\usepackage{mathptmx}
\usepackage{etoolbox}
\usepackage{multirow}
\usepackage{hyperref} 

\makeatletter
\def\@email#1#2{%
	\endgroup
	\patchcmd{\titleblock@produce}
	{\frontmatter@RRAPformat}
	{\frontmatter@RRAPformat{\produce@RRAP{*#1\href{mailto:#2}{#2}}}\frontmatter@RRAPformat}
	{}{}
}%
\makeatother

\graphicspath{{./Figures/}}
\newcolumntype{P}[1]{>{\centering\arraybackslash}p{#1}}

\begin{document}
	
	\title[]{Understanding In-Chamber Plasma Behavior Using a Dimensionally Scaled Gridded Ion Thruster in Three-Dimensional Kinetic Particle-in-Cell Simulations}
	
	\author{Gyuha Lim}
	\email{gyuhal2@illinois.edu}
	\author{Deborah Levin}
	\affiliation{Department of Aerospace Engineering, University of Illinois, Urbana-Champaign.}
	
	\date{\today}
	
	\begin{abstract}
		We investigate facility effects on a reduced-scale gridded ion thruster plume using a fully kinetic, three-dimensional Particle-in-Cell/Monte Carlo Collision (PIC-MCC) solver coupled with a Direct Simulation Monte Carlo (DSMC) neutral background. This approach enables detailed examination of key plasma processes governing beam neutralization and wall interactions under ground-test conditions. We find that inelastic electron cooling is essential for achieving a physically consistent, neutralized beam. Increasing the background pressure enhances ion-neutral collisions, leading to more charge- and momentum-exchange events that reduce ion mean energies, broaden the beam, and increase sidewall losses. Including inelastic processes flattens the potential, sustains quasi-neutrality, and preserves beam collimation farther downstream. Single-particle trajectory analyses show that primary electrons undergo mixed escape and temporary trapping, while low energy post-inelastic electrons remain confined, sustaining the neutralization cloud. Sheath diagnostics reveal that at the beam dump, classical Child-Langmuir and Hutchinson models underpredict the sheath length due to residual electrons, while near the sidewall, the sheath is truncated by beam-sheath interference within the compact domain. Current-flow analysis indicates that higher background pressure conditions yield lower beam energies and increased sidewall currents.
	\end{abstract}
	
	\maketitle
	
	\section{Introduction}
	Electric propulsion (EP) is increasingly central to modern spacecraft architecture, particularly for small satellites and constellation missions where high efficiency and long operational lifetimes are essential~\cite{GoebelKatzMikellides2025,LevchenkoNatureComm2018}. Among EP options, gridded ion engines (GITs) are recognized as a leading architecture capable of scaling to high power densities with high specific impulse~\cite{FosterAIAA2000}. Ground-based testing of GITs is indispensable for development and qualification. However, elevated neutral pressures, finite chamber dimensions, grounded walls, and sputtered material from chamber walls can distort the plasma plume and electric field, biasing key performance metrics such as thrust, specific impulse, beam divergence, and lifetime~\cite{FosterPoP2024}. Understanding and mitigating these facility effects is especially critical for the next generation of high-power-density thrusters.
	
	Two main classes of plasma plume simulation models have been widely employed. In the first approach, ions are tracked as macroparticles while electrons are treated as a massless fluid obeying the Boltzmann relation. Although computationally efficient, Hu and Wang showed the electrons in a plasma plume are non-equilibrium and the electron temperature is anisotropic. Thus, the electron fluid method neglects the anisotropic and non-equilibrium character of electrons and does not resolve the beam neutralization region, which can lead to overestimation of plume potentials and incorrect electron cooling~\cite{HuIEEE2015, WangPoP2019}. The second type of models are fully kinetic approach because they model both ions and electrons to investigate plasma plume and neutralization processes\cite{WangIEEE2012, HuIEEE2015, HuPSST2020, WangIEEE2015, LiPSST2019, WangPoP2019, NuwalIEEE2020, JambunathanIEEE2020, NishiiJPP2023, Usui2013, JambunathanJCP2018, BriedaAIAA2005, BriedaIEEE2018,JambunathanIEEE2020-CCE, NishiiPSST2023, GuaitaPSST2025}.    
	Table~\ref{Table:past} summarizes prior fully kinetic studies of beam neutralization and highlights the main characteristics of the present simulation framework.

	\begin{table*}[htbp]
		\centering
		\caption{Previous studies of the gridded ion thruster plume using a fully kinetic PIC approach.}
		\label{Table:past}
		\begin{tabular}{lccccc}
			\hline
			Author$^{\mathrm{Ref}}$  & Neutralizer  & Species & Ion-neutral & Electron-neutral & Geometry \\
			& position &  & collision & collision &  \\
			\hline
			Hu and Wang\cite{HuIEEE2015} & Co-located & Proton   & No & No & 2D In-space \\		
			Wang et al.\cite{WangIEEE2012} & Co-located & Proton   & No & No & 2D In-space\\
			Hu et al. \cite{HuPSST2020} & Co-located & Proton   & CEX$^{b}$ & No & 2D In-space \\
			Wang et al. \cite{WangIEEE2015} & Co-located & Argon   & CEX$^{b}$ & No & 3D In-chamber \\
			Li et al.\cite{LiPSST2019} & Co-located & Proton   & No & No & 2D In-space \\
			Wang and Yuan\cite{WangPoP2019} & Co-located & Xenon   &No  & No & 2D In-space \\
			Nuwal et al. \cite{NuwalIEEE2020} & Co-located & Xenon   & CEX/MEX$^{c}$ & No & 3D In-space w/SAP$^{a}$ \\
			Jabunathan and Levin\cite{JambunathanIEEE2020} & Co-located & Xenon   & CEX/MEX$^{c}$ & No & 3D In-space w/SAP$^{a}$ \\
			Nishii and Levin\cite{NishiiJPP2023} & Co-located & Xenon   & CEX/MEX$^{c}$ & No & 3D In-chamber \\
			Usui et al. \cite{Usui2013} & External & Proton   & No & No & 3D In-space \\
			Jambunathan and Levin\cite{JambunathanJCP2018} & External & Xenon   & No & No & 3D In-space \\
			Brieda and Wang \cite{BriedaAIAA2005} & External & Xenon   & No & No & 3D In-space \\		
			Brieda \cite{BriedaIEEE2018} & Internal & Oxygen &   No & No & 2D In-space \\
			Jabunathan and Levin\cite{JambunathanIEEE2020-CCE} & External & Xenon   & No & No & 3D In-space \\
			Nishii and Levin\cite{NishiiPSST2023} & External & Xenon   & CEX/MEX$^{c}$ & No & 3D In-chamber \\
			Guaita et al.\cite{GuaitaPSST2025} & External & Xenon & CEX/MEX$^{b}$ & elastic/inelastic & 2D In-chamber \\
			This study & External & Xenon   & CEX/MEX$^{b}$ & elastic/inelastic & 3D In-chamber \\
			\hline
		\end{tabular}

		\begin{flushleft}
			\footnotesize
			$^{a}$ SAP stands for a solar array panel geometry. \\
			$^{b}$ Calculated by Monte Carlo Collisions calculation (MCC).\\
			$^{c}$ Calculated by Direct Simulation Monte Carlo calculation (DSMC).\\
		\end{flushleft}
	\end{table*}

	Using a kinetic approach, Nishii and Levin performed three-dimensional Particle-In-Cell Direct Simulation Monte Carlo (PIC-DSMC) simulations to compare a low-density gridded ion thruster under in ground-test versus in-space conditions~\cite{NishiiPSST2023}. Their results showed that elevated neutral pressure enhances charge-exchange (CEX) and momentum-exchange (MEX) collisions, leading to the formation of ion sheaths near the chamber walls. They also demonstrated that the position of the neutralizer strongly influences the formation of a plasma bridge between the neutralizer and the ion beam, underscoring the importance of modeling neutralizer placement and wall boundary conditions when investigating beam neutralization.

	More recently,  experiments by Topham et al.~\cite{TophamIEPC2024,TophamSciTech2025} further characterized facility effects by measuring beam properties, neutralizer coupling, and near-wall plasma properties in gridded ion thrusters. While the PIC-DSMC approach of Nishii and Levin~\cite{NishiiPSST2023} accurately captures rarefied plasma behavior, it becomes computationally prohibitive when applied to plasma densities one to two orders of magnitude higher, consistent with values used in recent gridded-ion-thruster experiments.\cite{TophamIEPC2024, TophamSciTech2025} In addition, the continued use of the more computationally expensive DSMC collision modeling, instead of MCC, to model neutral-ion collisions is no longer required at higher plasma density conditions. Finally, simply increasing the plasma density within such simplified geometries leads to the formation of a virtual anode near the thruster exit, indicating that a direct scaling of the earlier low-density thruster models cannot reproduce realistic potential distributions under facility-relevant conditions. 
	
	The work of 
	Guaita et al.~\cite{GuaitaPSST2025} demonstrated the  importance of including electron-xenon collisions in a kinetic framework to accurately capture beam neutralization dynamics under ground facility conditions.  Similarly, we found in our earlier DSMC-informed PIC-MCC simulations of a dimensionally scaled version of the NSTAR Little Brother configuration that inclusion of electron-neutral interactions resolves the formation of the high-potential (virtual-anode) formation near the thruster exit.\cite{LimIEPC2025} Furthermore, the use of a dimensionally scaled model allows us to study the key 
	key plasma processes and scaling behavior prior to pursuing computationally intensive full-scale simulations. Whereas our previous work\cite{LimIEPC2025} primarily examined the transient behavior and initial validation of this reduced-scale model, the present study provides a more detailed, physics-based analysis of the gridded-ion-thruster plume under ground-testing conditions, with particular attention to the influence of background pressure on plume dynamics and beam neutralization, building upon the same computational framework.   Examination of the last two rows of Table~\ref{Table:past} also shows that the simulations results presented here are the first for a fully three dimensional configuration.

	The remainder of the paper is as follows: the DSMC neutral solution is first presented, followed by analyses of electron-xenon inelastic collisions and the influence of background pressure on plume expansion. The subsequent sections examine beam characteristics and single-electron trajectory analyses that clarify beam-neutralization processes. Finally, the sheath structures and overall current-flow pathways within the chamber are discussed. Together, these results establish a comprehensive picture of in-chamber plasma behavior and provide a framework for future full-scale kinetic simulations.

	\section{Computational Methods}
	
	Three–dimensional particle–based simulations were performed using the CUDA-based Hybrid Approach for Octree Simulation (CHAOS) framework~\cite{JambunathanJCP2018}. The overall methodology combines a Direct Simulation Monte Carlo (DSMC) module for neutral flow, a Particle–In–Cell with Monte Carlo Collisions (PIC–MCC) module for charged species, and a hybrid mapping strategy that couples the two. The subsections below summarize the main algorithmic steps.
	
	\subsection{Direct Simulation Monte Carlo}
	The DSMC module~\cite{BirdDSMC1994} simulates the neutral xenon flow by representing real particles as superparticles of weight $\mathrm{Fnum}$. Each DSMC time step $\Delta t_n$ consists of two main stages: (i) sampling and executing neutral–neutral collisions, and (ii) ballistic particle advancement. In each cell, the neutral–neutral collision frequency is calculated as  
	\begin{equation}
		\nu_{gg} = n_g \,\sigma_{gg}\,\bar{c}_{gg},
	\end{equation}
	where $n_g$ is the local neutral number density, $\sigma_{gg}$ is the elastic collision cross section, and $\bar{c}_{gg}$ is the mean relative speed. Collisions are executed as elastic scatterings in the center-of-mass frame while conserving momentum and kinetic energy. After the collision stage, all neutral superparticles are advanced ballistically over $\Delta t_n$. Interactions with boundaries result in either absorption or reflection according to the specified surface accommodation coefficient.  
	
	To ensure physical fidelity, the DSMC time step satisfies  
	\begin{equation}
		\Delta t_n < 0.1\,\frac{\lambda_{\rm mfp}}{\bar{c}_{gg}},
	\end{equation}
	where the local mean free path is $\lambda_{\rm mfp} = 1/(n_g\,\sigma_{gg})$.
	
	\subsection{Particle–In–Cell with Monte Carlo Collisions}
	\begin{figure}[htbp]
		\centering
		\includegraphics[width=0.5\textwidth]{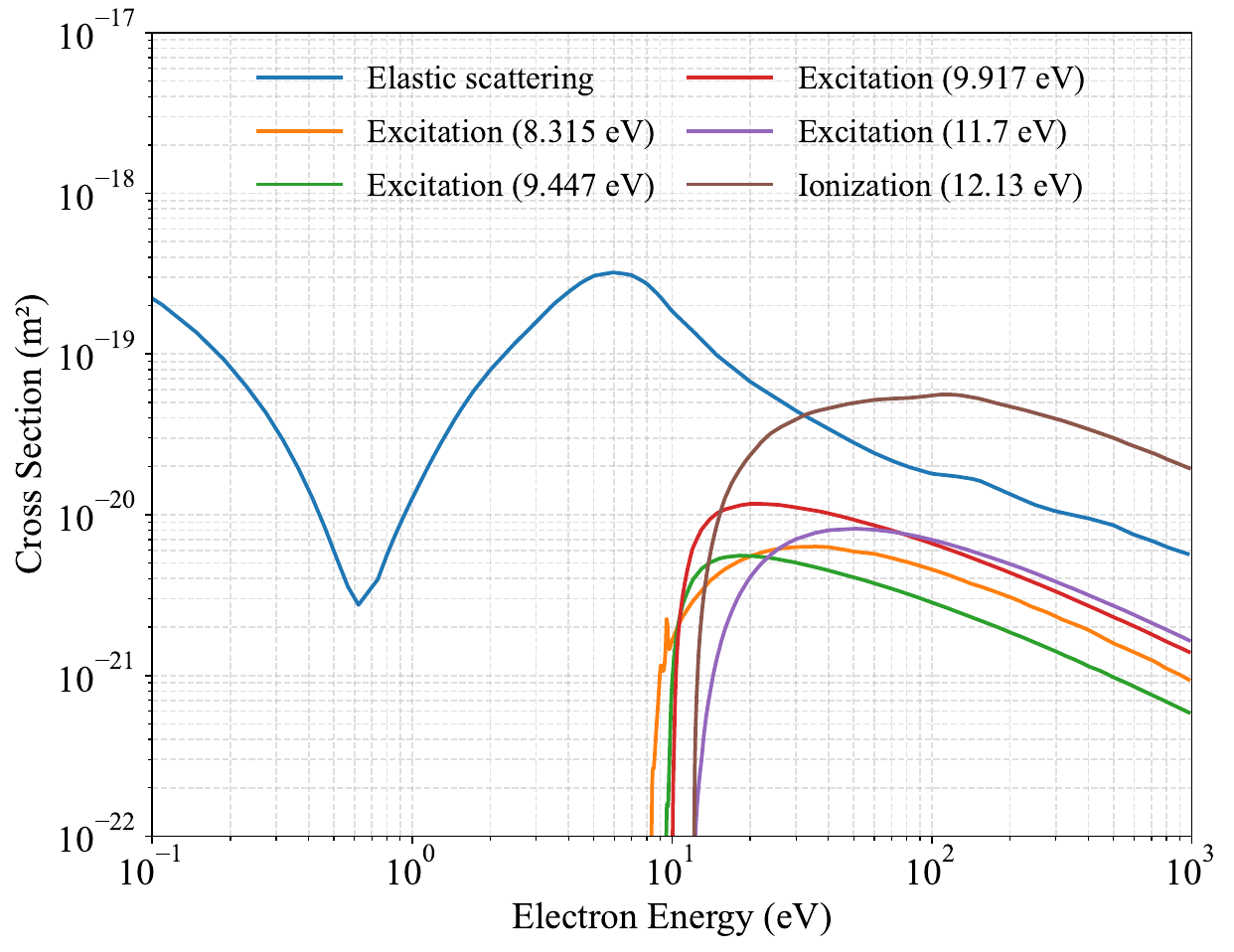}
		\caption{Electron–xenon collision cross sections (Biagi v7.1) from LXCat~\cite{LXCat}. }
		\label{Fig:cross}
	\end{figure}

	The Particle-In-Cell Monte Carlo Collisions (PIC–MCC) module advances ions and electrons under self-consistently computed electric fields while modeling their collisions with a prescribed neutral background. At each time step (typically picoseconds to nanoseconds), charge is deposited from particles onto the CHAOS Electrical Forest-of-Trees (E-FOT) grid, and Poisson’s equation is solved to obtain the electric potential:  
	\begin{equation}
		\nabla^2\phi = -\frac{\rho}{\varepsilon_0},
		\quad 
		\mathbf{E} = -\nabla\phi.
	\end{equation}
	Particles are then pushed in phase space using the computed electric field.
	
	Collisions between ion-neutral and electron-neutral are modeled using with the null-collision method~\cite{Nanbu1997}. For each particle, the probability of a collision in $\Delta t$ is  
	\begin{equation}
		P = 1 - \exp(-\nu\,\Delta t),
		\quad 
		\nu = n_g \sum_k \sigma_k\,v,
	\end{equation}
	where $n_g$ is the neutral density, $\{\sigma_k\}$ are the cross sections for each process, and $v$ is the particle speed.  
	
	This study includes MEX and CEX collisions for ions, as well as electron–neutral elastic, ionization, and multiple excitation processes. CEX produces slow ions and fast neutrals via electron transfer, whereas MEX redistributes momentum without changing charge state. Inelastic electron–neutral processes have been shown to play an important role in beam neutralization~\cite{GuaitaPSST2025}. Figure~\ref{Fig:cross} shows the electron–xenon collision cross sections obtained from the LXCat database~\cite{LXCat} that were used in this work.

	\subsection{PIC–MCC with a DSMC informed neutral background}
	
	Ground tests of gridded ion thrusters exhibit spatially nonuniform neutral distributions~\cite{SoulasJPP2011}. Although a fully coupled PIC–DSMC approach could resolve these distributions directly, it is computationally prohibitive and often unnecessary. In this work, we adopt a hybrid strategy: a cold-flow DSMC simulation is first performed to generate the neutral number density and temperature fields, which are then mapped into the PIC–MCC solver as a spatially varying background.  Table~\ref{Table:numerical_parameter} summarizes the numerical parameters used in two simulations for all cases. 
	
	\begin{figure*}[tbp]
		\centering
		\includegraphics[width=.95\textwidth]{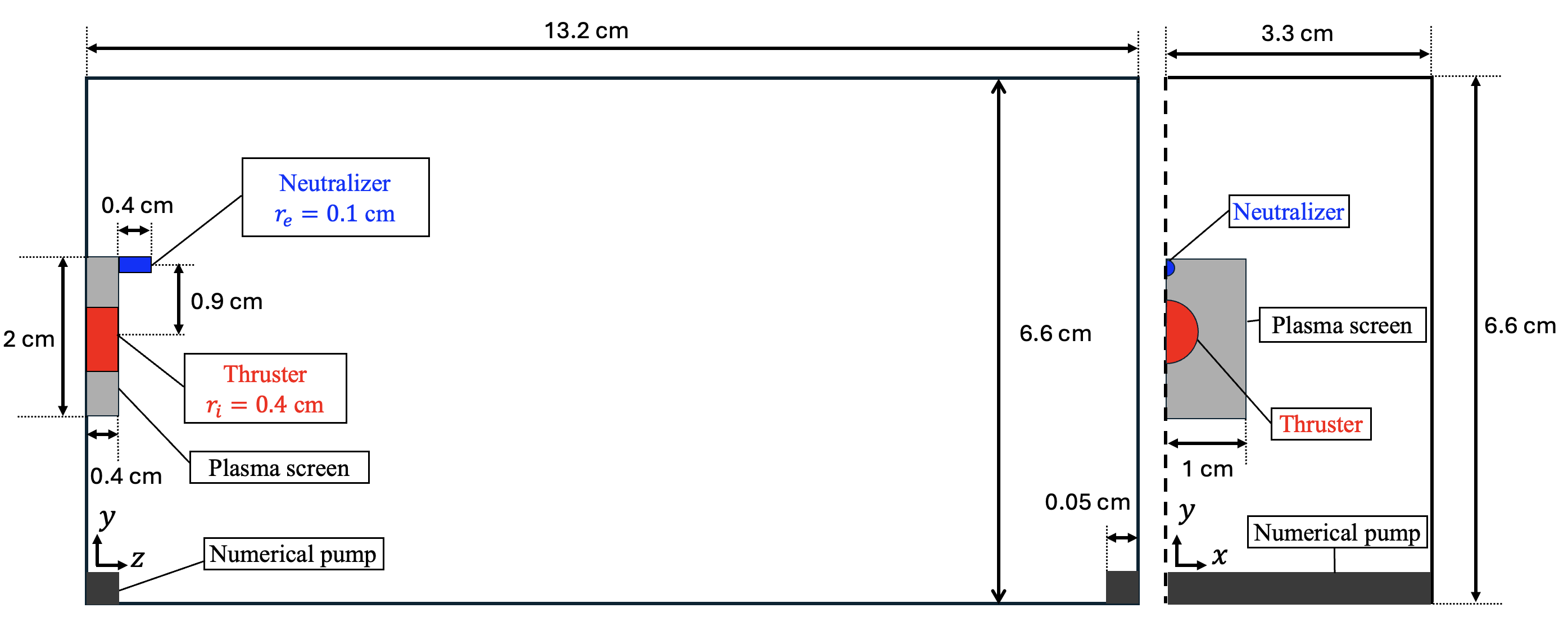}	
		\caption{Simulation schematic of the NSTAR–Little Little Brother setup: $yz$-plane cross-section (left) and $xy$-plane cross-section (right). Key dimensions are shown in centimeters. $r_i$ and $r_e$ refer to the radii of the ion beam thruster and neutralizer, respectively. Figures are not to scale.}
		\label{Fig:setup}
	\end{figure*}

	This approach retains the essential nonuniformity of the neutral environment while significantly reducing the computational cost compared with a fully coupled PIC–DSMC simulation. Both memory usage and runtime are improved, while collision modeling in the vacuum chamber is much more realistic than assuming a uniform neutral background.  A comparison among uniform versus spatially varying DSMC backgrounds used in PIC-MCC simulations with a fully coupled PIC-DSMC will be presented in future research~\cite{LimSciTech2026}.

\begin{table}[htbp]
	\caption{Numerical parameters used in this study}
	\label{Table:numerical_parameter}
	\begin{ruledtabular}
		\begin{tabular}{lcc}
			Simulation              & \textbf{DSMC}      & \textbf{PIC--MCC}        \\
			\hline
			Species                 & \textbf{Xe}        & \textbf{Xe$^+$/e$^-$}    \\
			$\Delta t$ (s)          & $2\times10^{-7}$   & $1\times10^{-11}$        \\
			Fnum$^{*}$              & $2\times10^{6}$    & 1000                     \\
			$N_{\mathrm{total}}^{**}$ 
			& $9.2\times10^{6}$  & $5.4\times10^{6} / 4.2\times10^{6}$ \\
			Timesteps               & $1.2\times10^{6}$  & $4.4\times10^{6}$        \\
		\end{tabular}
	\end{ruledtabular}
	
	\vspace{1ex}
	\begin{flushleft}
		\footnotesize
		$^{*}$ Fnum refers to the number of real particles per each computational particle. \\
		$^{**}$ $N_{\mathrm{total}}$ refers to the total number of particles for each species
		in the whole domain (values shown for case~1A). \\
	\end{flushleft}
\end{table}

	\section{Simulation Setup}
\begin{table}[t]  
	\caption{Inlet conditions at the thruster and neutralizer exits}
	\label{Table:inlet}
	\begin{ruledtabular}
		\begin{tabular}{lcccc}
			& \multicolumn{2}{c}{Thruster} & \multicolumn{2}{c}{Neutralizer} \\
			\textrm{Parameter}
			& Xe
			& Xe$^+$
			& Xe
			& e$^-$ \\
			\hline
			Inlet distribution
			& drift Maxwellian
			& Gaussian beam (11$^\circ$)
			& drift Maxwellian
			& half Maxwellian \\
			Number density (m$^{-3}$)
			& $2.6\times10^{17}$
			& $2.3\times10^{15}$
			& $2.5\times10^{18}$
			& $1.0\times10^{16}$ \\
			Temperature (K)
			& 600
			& 600
			& 1500
			& 23212 (2 eV) \\
			Bulk velocity (m/s)
			& 10
			& 34300
			& 430
			& -- \\
			Current density (A/m$^2$)
			& --
			& 7.99
			& --
			& 379.0 \\
			Current (mA)
			& --
			& 0.40
			& --
			& 1.19 \\
		\end{tabular}
	\end{ruledtabular}
\end{table}

	In this study, we adopt the configuration of the NSTAR--Little Brother device (8~cm thruster diameter) recently described by Topham et al.~\cite{TophamIEPC2024, TophamSciTech2025}, itself a reduced-scale version of the original NSTAR thruster. Because direct simulation of the full chamber geometry is computationally expensive, we further downscale the system by an order of magnitude, creating the NSTAR--Little Little Brother device (0.8~cm thruster diameter). This miniature configuration enables us to investigate the fundamental physics of beam neutralization before modeling experiments at the NSTAR--Little Brother length scale. Figure~\ref{Fig:setup} shows a schematic of the NSTAR--Little Little Brother geometry. A symmetry condition is applied along the $x$-axis, while all other boundaries represent vacuum chamber walls at 300~K with an accommodation coefficient of unity for neutral in DSMC and charge-absorbing species for ions and electrons conditions in PIC--MCC. The thruster assembly (thruster, neutralizer, and plasma screen) absorbs all charged particles while reflecting neutrals specularly, and a numerical pump removes particles that enter its region.  
	
	Table~\ref{Table:inlet} summarizes the species properties at the thruster and neutralizer exits. Ion parameters were derived from the beam voltage and current reported by Topham et al.~\cite{TophamIEPC2024}. Neutral temperatures were set to 600~K at the thruster exit and 1500~K at the neutralizer exit, consistent with the reported values of 300\,$^\circ$C and 1100\,$^\circ$C by Soulas~\cite{SoulasJPP2011}. At the thruster exit, the neutral bulk velocity was modeled as a half-Maxwellian corresponding to 600~K, while the neutralizer flow was assumed to choke, producing a supersonic jet with an exit velocity of 440~m/s (Mach~1.08). Neutral number densities were then computed from the specified temperature, bulk velocity, and mass flow rate. The electron current was set to $I_e \approx 3\,I_i$, following Guaita et al.’s sensitivity analysis, which demonstrated minimal impact of $I_e$ on downstream results~\cite{GuaitaPSST2025}.  
	
	\begin{table}[hbtp]
		\centering
		\caption{Simulation case characteristics}
		\label{Table:case}
		\begin{tabular}{cccc}
			\hline
			\textbf{Case}       & Xe-Xe+  & Xe-e  & Pressure level \\
			\hline
			0A       &CEX, MEX\cite{ArakiIEEE2013}      & No collision      & $\sim20\mu \mathrm{Torr}$  \\
			1A       &CEX, MEX\cite{ArakiIEEE2013}      & elastic, excitation, ionization\cite{LXCat}      & $\sim20\mu \mathrm{Torr}$  \\
			1B       &CEX, MEX\cite{ArakiIEEE2013}      & elastic, excitation, ionization\cite{LXCat}      & $\sim40\mu \mathrm{Torr}$ \\
			\hline
		\end{tabular}
	\end{table}

	To construct the NSTAR--Little Little Brother model, dimensional scaling was applied while keeping intensive properties (temperature, density, velocity) identical to those of Little Brother~\cite{TophamIEPC2024, TophamSciTech2025}. Consequently, absolute currents shown in Table~\ref{Table:inlet} were reduced by a factor of 100, consistent with the 10$\times$ geometric downscaling. DSMC simulations were first performed under these baseline conditions fo 20~$\mu$Torr, and an additional case was examined by raising the background pressure by 20~$\mu$Torr through the introduction of uniform neutral density, in order to match experimental conditions and assess the effect of pressure variation. Subsequent PIC--MCC simulations were then carried out for both neutral background profiles. Table~\ref{Table:case} summarizes the simulation cases performed.

	\section{Results}
	\subsection{DSMC solution for neutral background}
	
	\begin{figure}[htbp]
		\centering
		\includegraphics[width=0.7\linewidth]{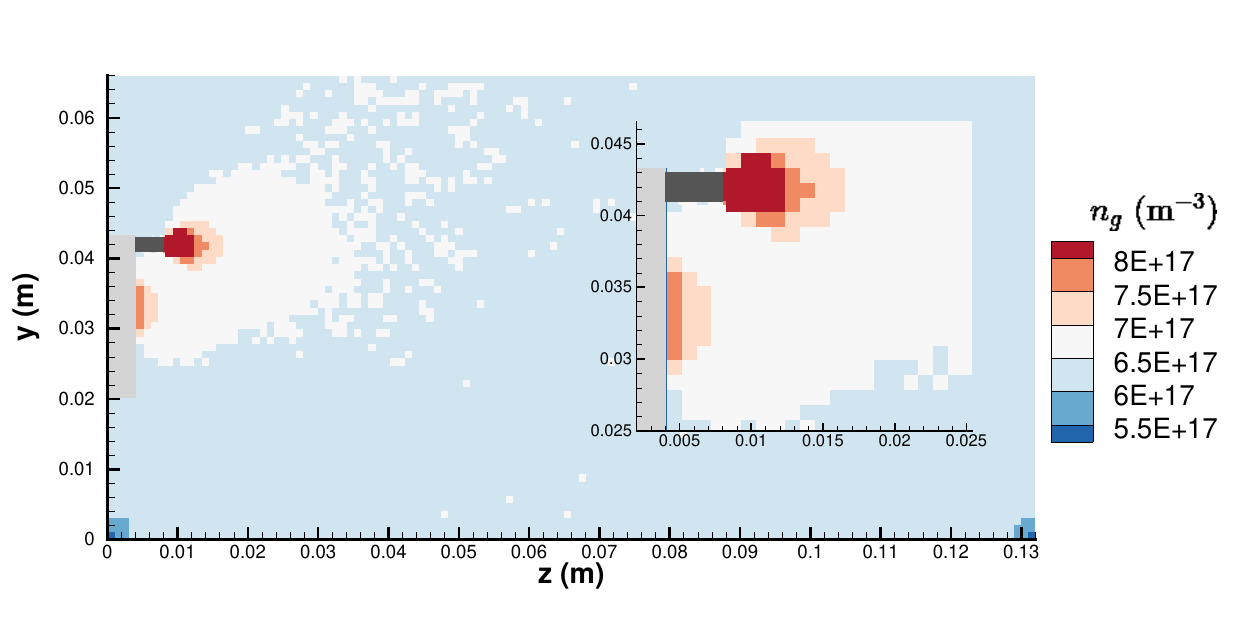}
		\includegraphics[width=0.7\linewidth]{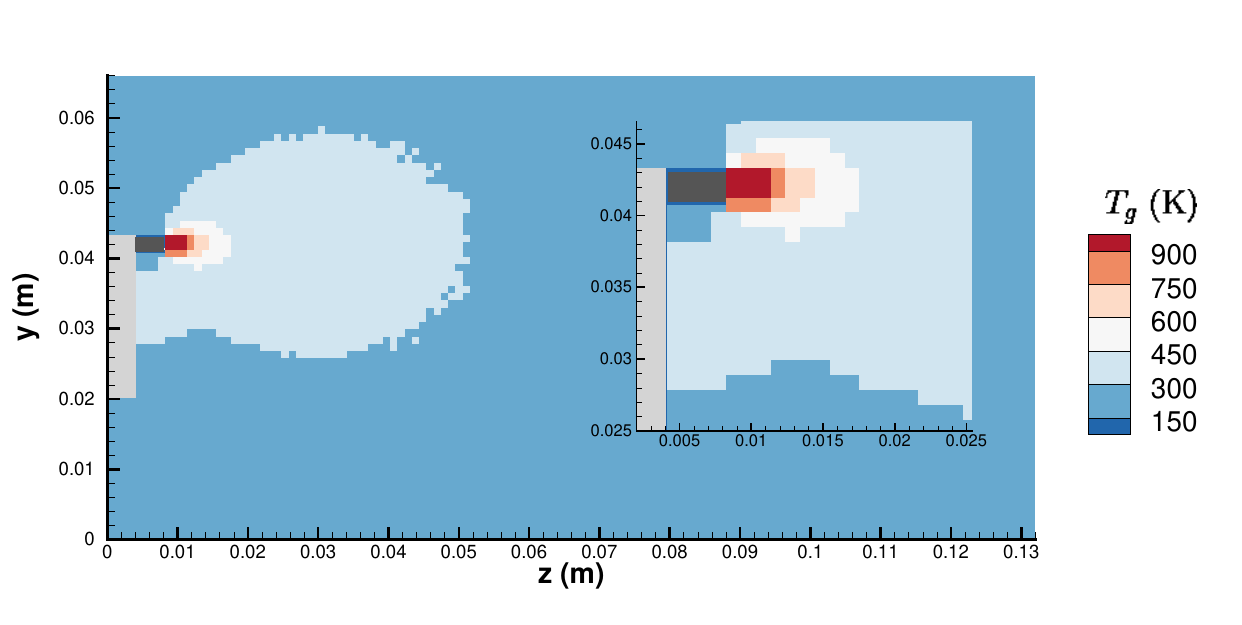}	
		\caption{Xenon neutral distributions in the $yz$--plane in the $x=0$ plane. Top: neutral number density $n_g$ (m$^{-3}$). Bottom: neutral temperature $T_g$ (K). 
			The dark gray and light gray areas indicate the plasma screen and neutralizer, respectively.}
		\label{Fig:neutral}
	\end{figure}

	We first carried out a neutral-only DSMC simulation to obtain the background xenon distribution. Figure~\ref{Fig:neutral} presents the neutral number density and temperature in the $yz$-plane at $x = 0$. In the far field, both quantities remain nearly uniform, reflecting wall-accommodated neutrals that remain at 300~K. In contrast, elevated density and temperature appear near the thruster and neutralizer exits, where fresh neutrals are injected into the chamber. This effect is most pronounced in front of the neutralizer, where locally heated neutrals increase the temperature above the background level due to hotter inflow neutrals from the neutralizer.

	\begin{figure}[htbp]
		\centering
		\includegraphics[width=0.7\linewidth]{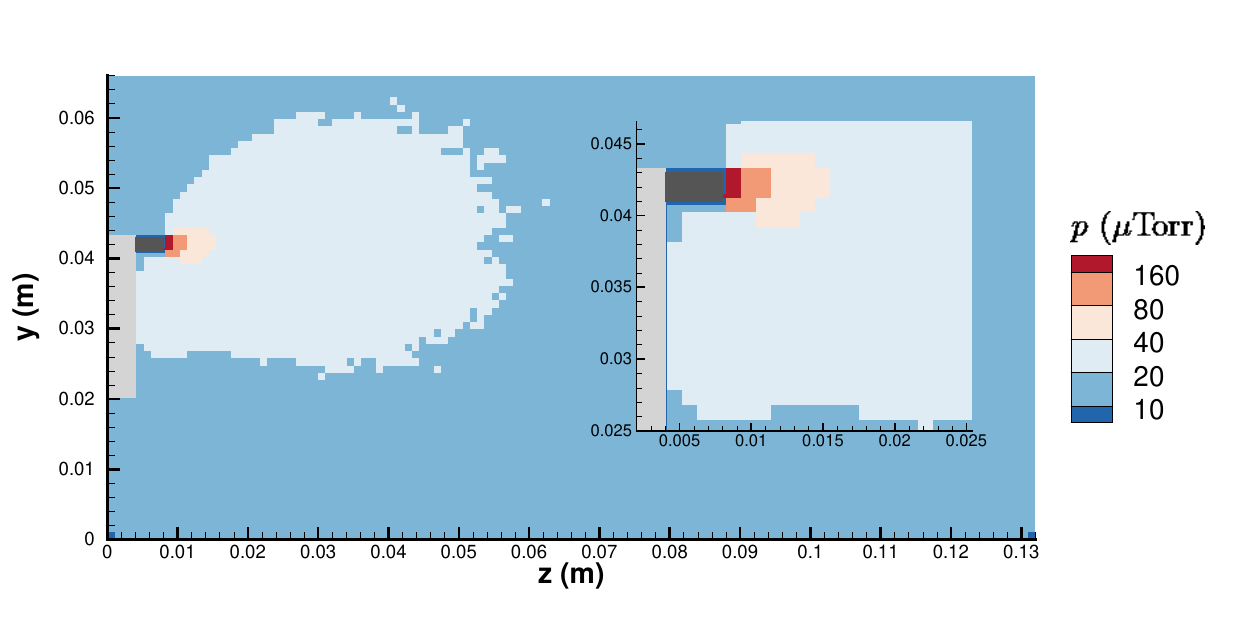}
		\includegraphics[width=0.7\linewidth]{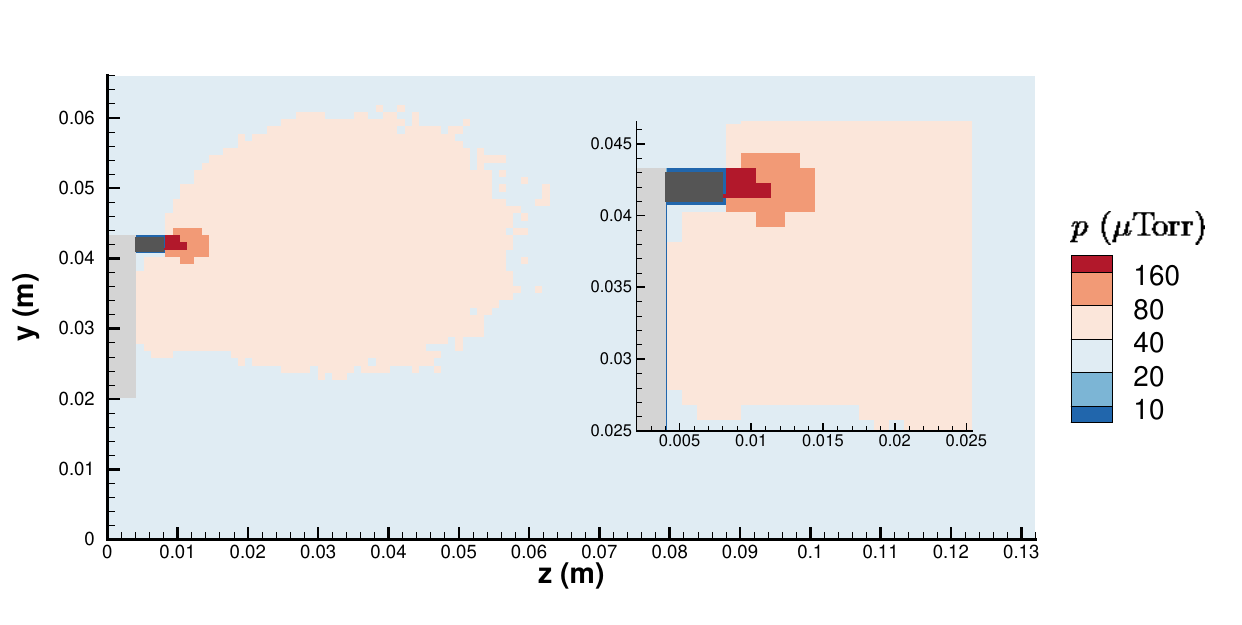}
		\caption{Xenon neutral pressure ($\mu\mathrm{Torr}$). Top: yz-plane at \(x=0\,\mathrm{m}\) from the DSMC solution, which is applied to case 0A and 1A. Bottom: the same plane with a uniform number-density offset applied so that \(p=(n+\Delta n)k_\mathrm{B}T\) matches the experimental pressure. This corrected background neutral field is used for case 1B. The light-gray structure is the plasma screen, and the dark-gray structure is the neutralizer.}
		\label{Fig:pressure}
	\end{figure}
	
	The corresponding pressure distribution from the DSMC results is shown in the top panel of Figure~\ref{Fig:pressure}, calculated as $p = n_g k_B T_g$. The representative chamber pressure, obtained at the thruster exit plane a short distance from the chamber wall, is approximately 20~$\mu$Torr. To match the experimental pressure level reported by Topham et al.~\cite{TophamIEPC2024} while maintaining the same temperature profiles, we added additional background xenon neutrals to increase the number density in the simulation. The resulting higher-pressure case is presented in the bottom panel of Figure~\ref{Fig:pressure}, which reproduces the measured pressure, 40~$\mu$Torr, and remains largely uniform throughout the chamber except for localized variations near the thruster and neutralizer exits. In particular, the peak pressure near the neutralizer exit reaches nearly 300~$\mu$Torr, which enhances Xe–Xe$^{+}$ CEX and MEX, and Xe–e collision frequencies in this region. These results highlight that the spatially nonuniform neutral background must be considered to accurately capture collision dynamics near the thruster exit. Both neutral background feeds for case 0A, 1A, and 1B.

	\subsection{Effect of inelastic collisions}
	
	\begin{figure}[htpb]
		\centering
		\includegraphics[width=0.45\linewidth]{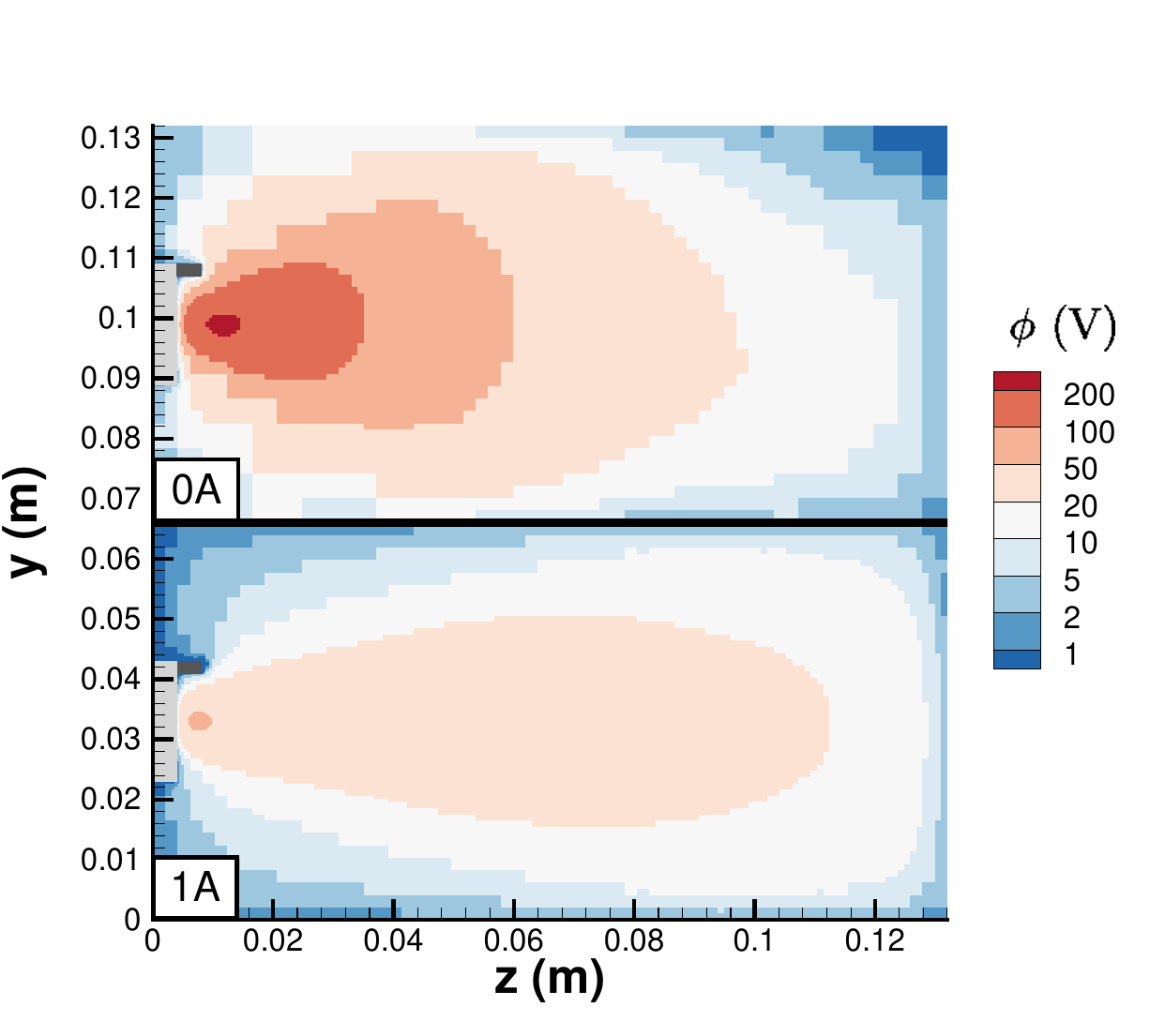}
		\includegraphics[width=0.45\linewidth]{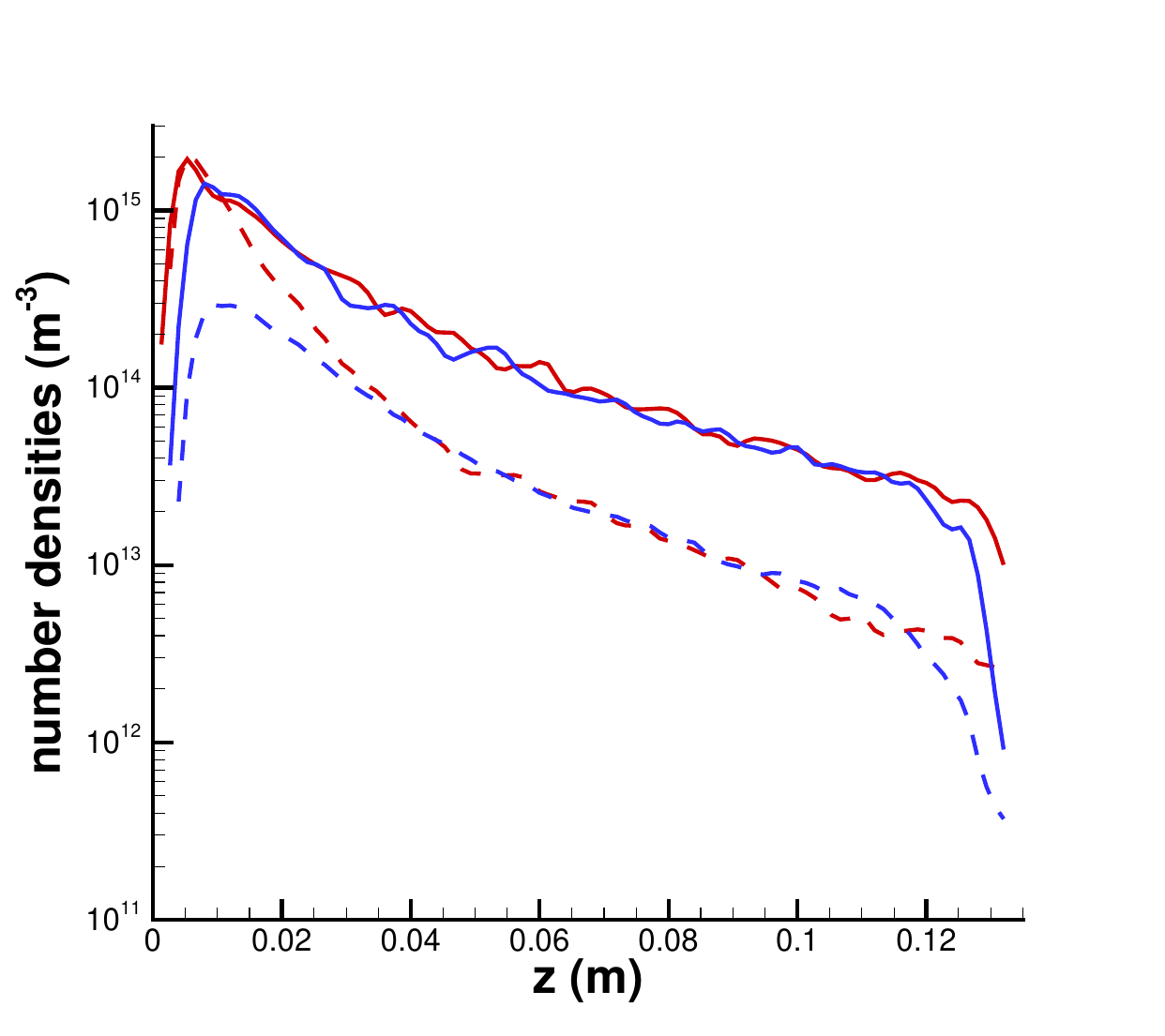}
		\caption{
			Left: Electric potential ($\phi$) contours in the $x=0$ plane for cases~0A (top) and~1A (bottom). Right: Axial number density profiles of ions (red) and electrons (blue) for cases~0A (dashed) and~1A (solid). 
			Data were post-processed using Tecplot smoothing to reduce numerical fluctuations.
		}
		\label{Fig:0Avs1A-Phi}		
		\vspace{0pt}
	\end{figure}
	
	Our previous study~\cite{LimIEPC2025} highlighted the importance of inelastic collisions in maintaining beam neutralization. 
	Further investigation confirms that, for the present configuration, inelastic collisions are indispensable for obtaining a physically consistent solution. 
	When inelastic processes are excluded as in case~0A, a strong potential gradient forms near the thruster exit and along the direction lateral to the plume due to the short distance to the grounded chamber walls. 
	This generates intense lateral electric fields that drive ions away from the beam axis, resulting in pronounced ion depletion and an unphysical charge distribution.

	\begin{figure}[htpb]
		\centering
		\includegraphics[width=0.7\linewidth]{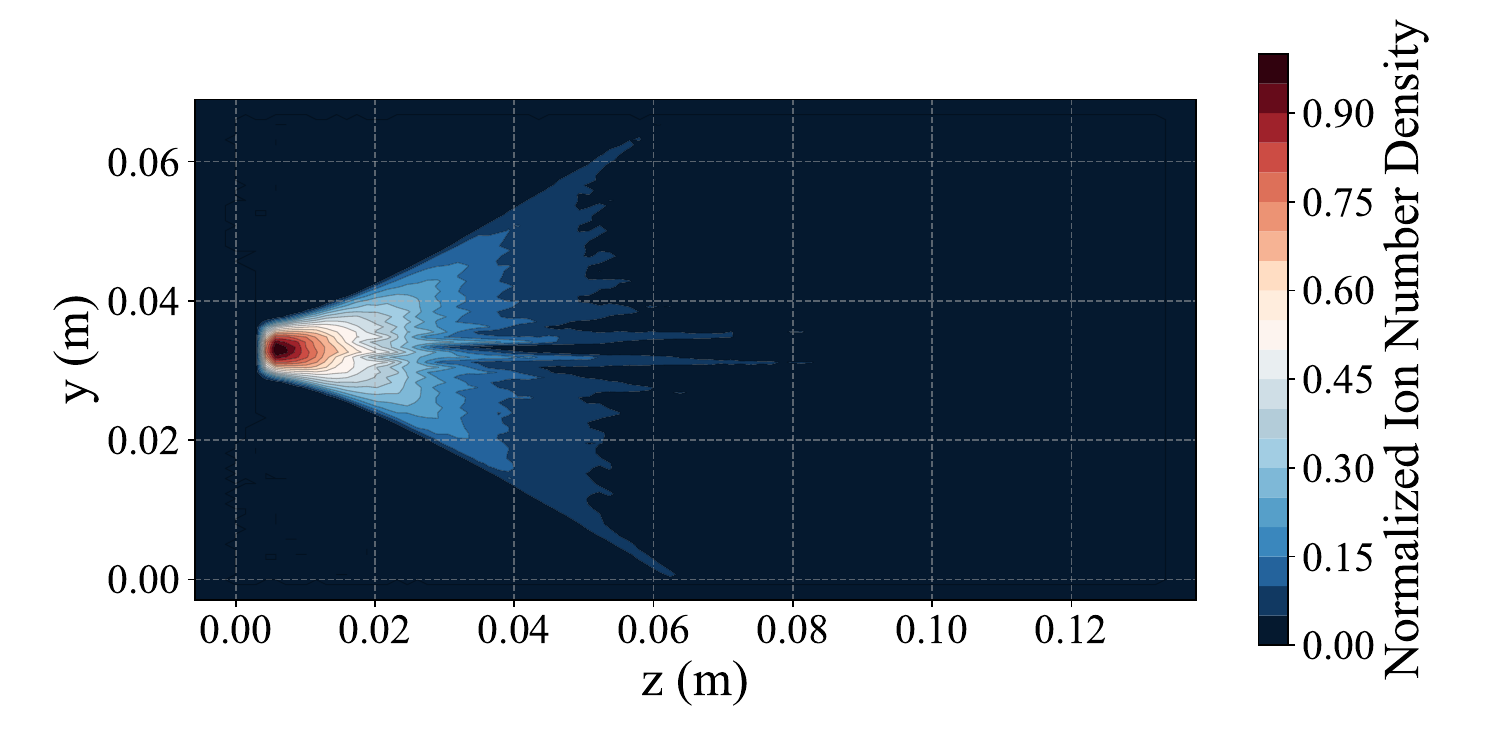}
		\includegraphics[width=0.7\linewidth]{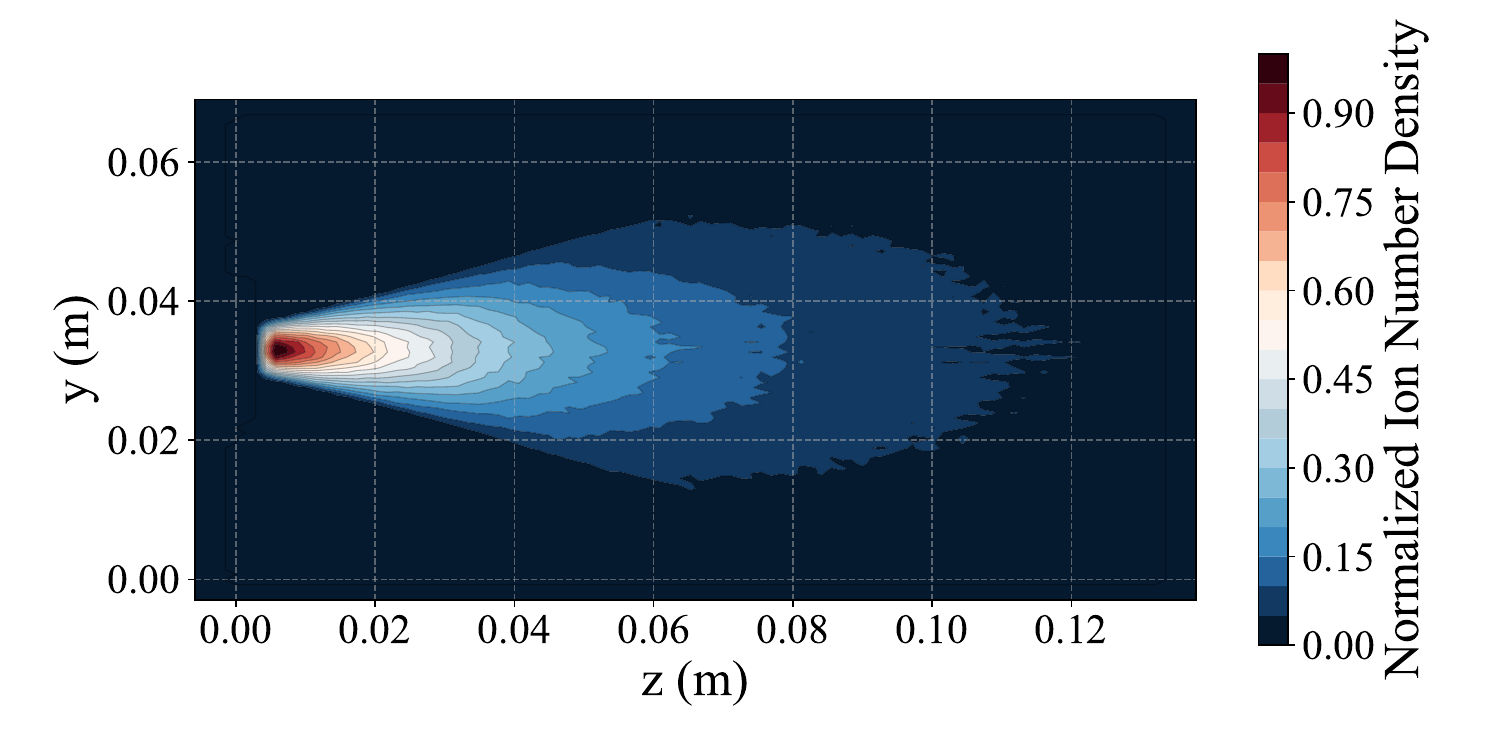}	
		\caption{
			Normalized ion number density contours in the $x=0$ plane for cases~0A (top) and~1A (bottom). 	}
		\label{Fig:0Avs1A-NDi}		
		\vspace{0pt}
	\end{figure}
	
	As shown in left panel of Figure~\ref{Fig:0Avs1A-Phi}, the electric potential in case~0A rises steeply up to nearly 200~V near the thruster exit. 
	The combination of this sharp potential gradient and the short distance to the grounded chamber walls produce an excessively strong lateral electric field. 
	This field drives ions away from the beam axis, resulting in a pronounced ion-depleted region along the centerline, as seen in Figure~\ref{Fig:0Avs1A-NDi}. 
	In contrast, when inelastic electron–neutral collisions are included as in case~1A, electron energy is dissipated through inelastic cooling, which lowers the electron temperature and reduces the local potential. 
	Consequently, the electric field becomes more moderate, leading to a physically realistic beam structure where ions remain concentrated near the centerline, forming a continuous and stable ion plume.

	Right panel of Figure~\ref{Fig:0Avs1A-Phi} provides a quantitative comparison of the ion and electron number density profiles along the thruster centerline. 
	As seen in the line plot, the ion density of case~0A drops sharply near the thruster exit, reflecting the strong divergence that expels ions from the beam core. At the same location, case~0A also exhibits pronounced electron depletion due to the absence of inelastic energy-loss processes, while case~1A maintains a higher electron population through collisional cooling. Further downstream, the overall plasma density remains higher in case~1A, indicating more effective beam neutralization and sustained quasi-neutrality compared with case~0A. 
	These findings reaffirm the essential role of inelastic collisions in stabilizing the charge distribution and producing physically valid plasma behavior. 
	Therefore, the subsequent analyses focus on the more representative cases~1A and~1B, which employ time-averaged over $4~\mu$s to reduce statistical noise and capture steady-state characteristics more clearly.

	\subsection{Effect of Pressure levels}
	\begin{figure}[htbp]
		\centering
		\includegraphics[width=0.6\linewidth]{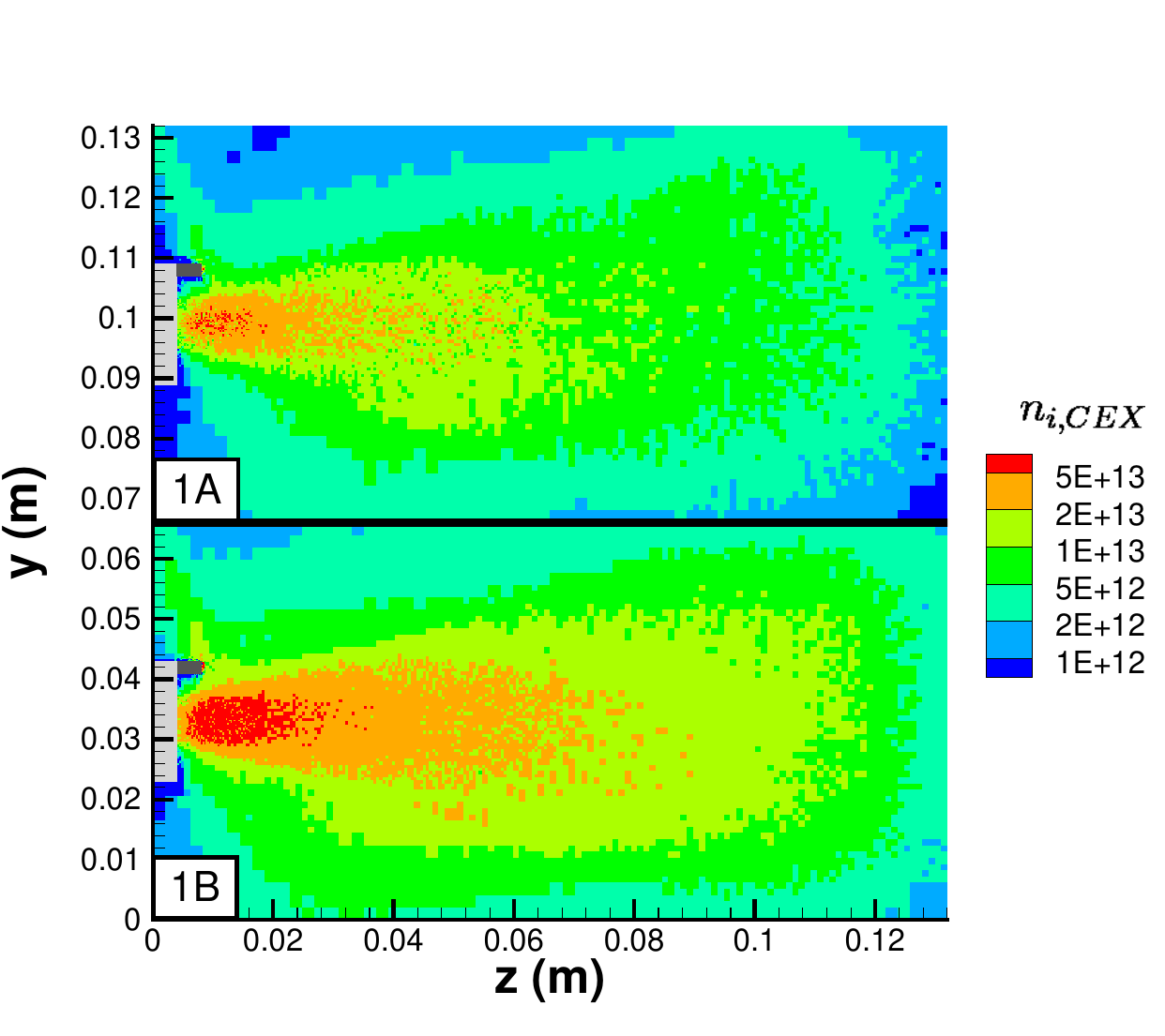}
		\caption{ CEX ion number density (ions/$\mathrm{m}^3$) contours in the $x=0$ plane for cases~1A (top) and~1B (bottom).}
		\label{Fig:1A1B-CEX}
	\end{figure}

	Collisional processes play a crucial role in determining the plume behavior under different pressure conditions. As the background pressure increases, CEX collisions become more frequent, leading to enhanced production of slow ions and increased charge neutrality within the beam. As seen in Figure~\ref{Fig:1A1B-CEX}, the higher-pressure case~1B exhibits more pronounced CEX ion generation, particularly near the thruster exit, resulting in a more balanced local charge distribution.
		\begin{figure}[htbp]
		\centering
		\includegraphics[width=0.6\linewidth]{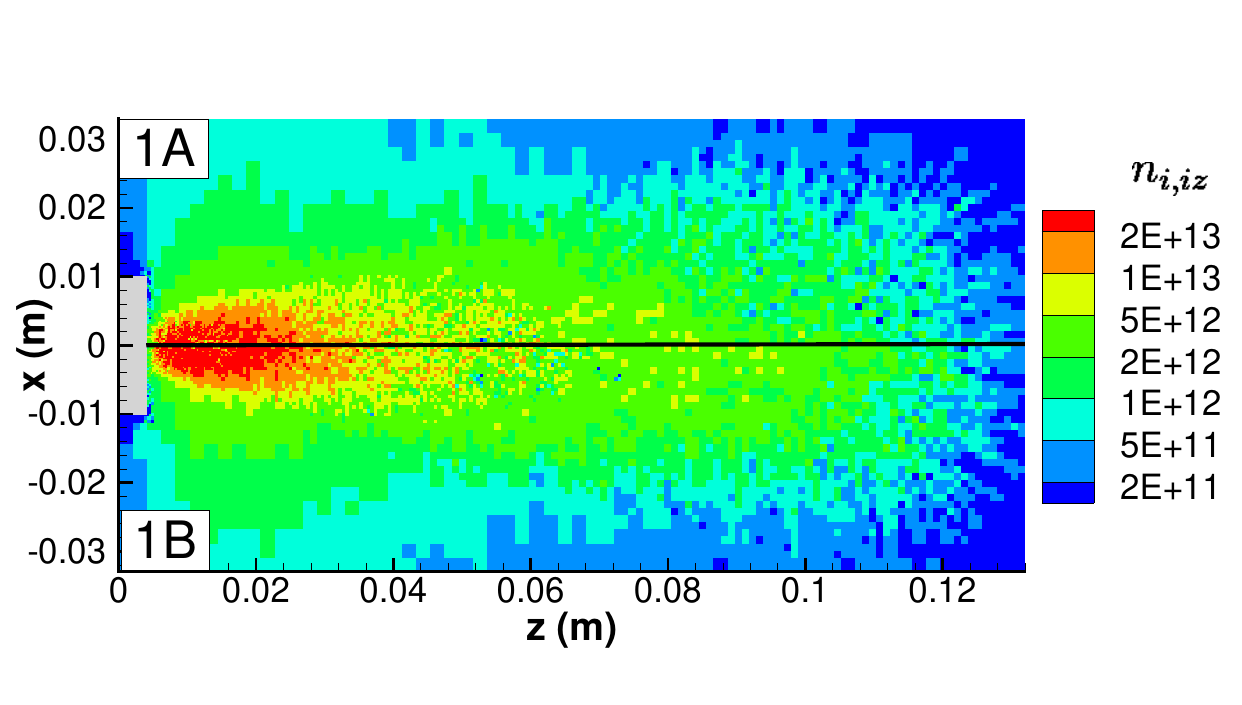}
		\caption{Ionization induced ion number density (ions/$\mathrm{m}^3$) contours at $y=0.033~\mathrm{m}$ for cases~1A (top) and~1B (bottom).}
		\label{Fig:1A1B-iz}
	\end{figure}
	
	Case~1B also exhibits slightly higher ionization-induced ion number densities near the thruster exit, as shown in Figure~\ref{Fig:1A1B-iz}. However, the difference in the number density of ions formed by ionization collisions between the two cases remains small because the elevated pressure enhances collisional cooling, leading to a lower electron temperature in case~1B. More frequent electron–neutral collisions also result in greater energy loss, and the reduced potential gradient in the plume further limits electron acceleration. Consequently, although more neutrals are available for impact ionization at higher pressure, the reduced electron energy suppresses inelastic collision rates, resulting in only a marginal overall increase in ionization-induced ion number density.

\begin{figure}[htbp]
		\centering
		\includegraphics[width=0.45\linewidth]{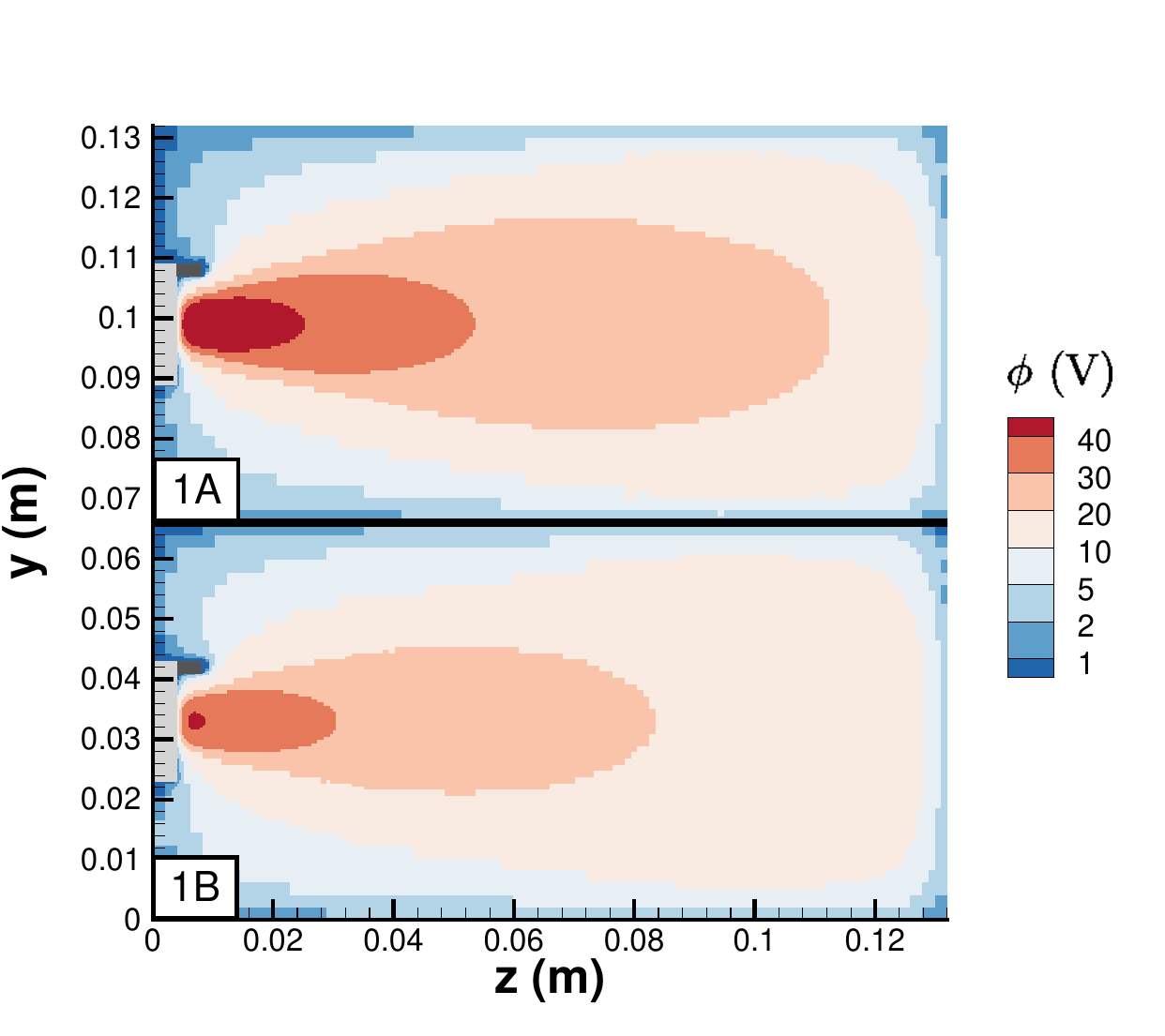}
		\includegraphics[width=0.45\linewidth]{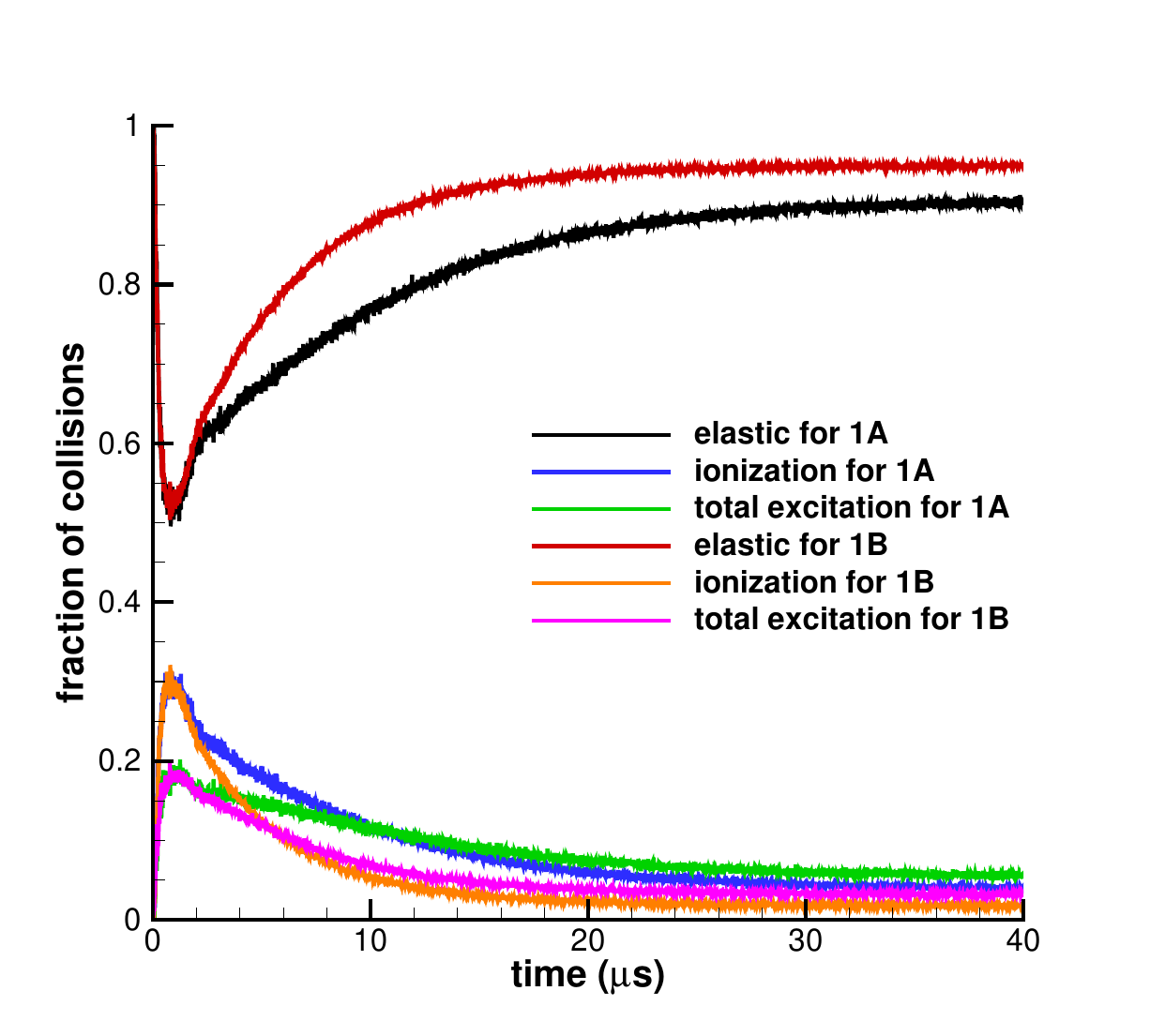}		
		\caption{		Left: Electric potential contours in the $x=0$ plane for cases~1A (top) and~1B (bottom). Right: Electron neutral collision fraction over time for cases 1A and 1B, averaged over the simulation domain. }
		\label{Fig:1A1B-Phi}
	\end{figure}

	Left panel of Figure~\ref{Fig:1A1B-Phi} compares the electric potential contours between cases~1A and~1B. 
	At higher background pressure (case~1B), the potential field becomes more flattened with lower peak electric potential near the thruster exit. 
	This behavior results from more frequent ion–neutral and electron–neutral collisions. In particular, the increased number of CEX ions in case~1B reduces the effective ion beam velocity, leading to broader and more uniform distributions of ion and electron number densities inside the chamber. 
	
	Right panel of Figure~\ref{Fig:1A1B-Phi} illustrates the time evolution of the electron–neutral collision fraction for both cases. 
	The higher-pressure case~1B exhibits a significantly larger number of inelastic collisions, 
	which accelerates the overall relaxation process of the plasma. 
	Frequent collisions enhance electron energy loss and thermal equilibration, allowing the plasma to reach steady state more rapidly than in case~1A. 
	However, the effect of increased background pressure is primarily seen to cause as faster temporal relaxation rather than a fundamental change in the steady-state potential structure.

	\begin{figure}[htpb]
		\centering
		\centering
		\includegraphics[width=0.45\linewidth]{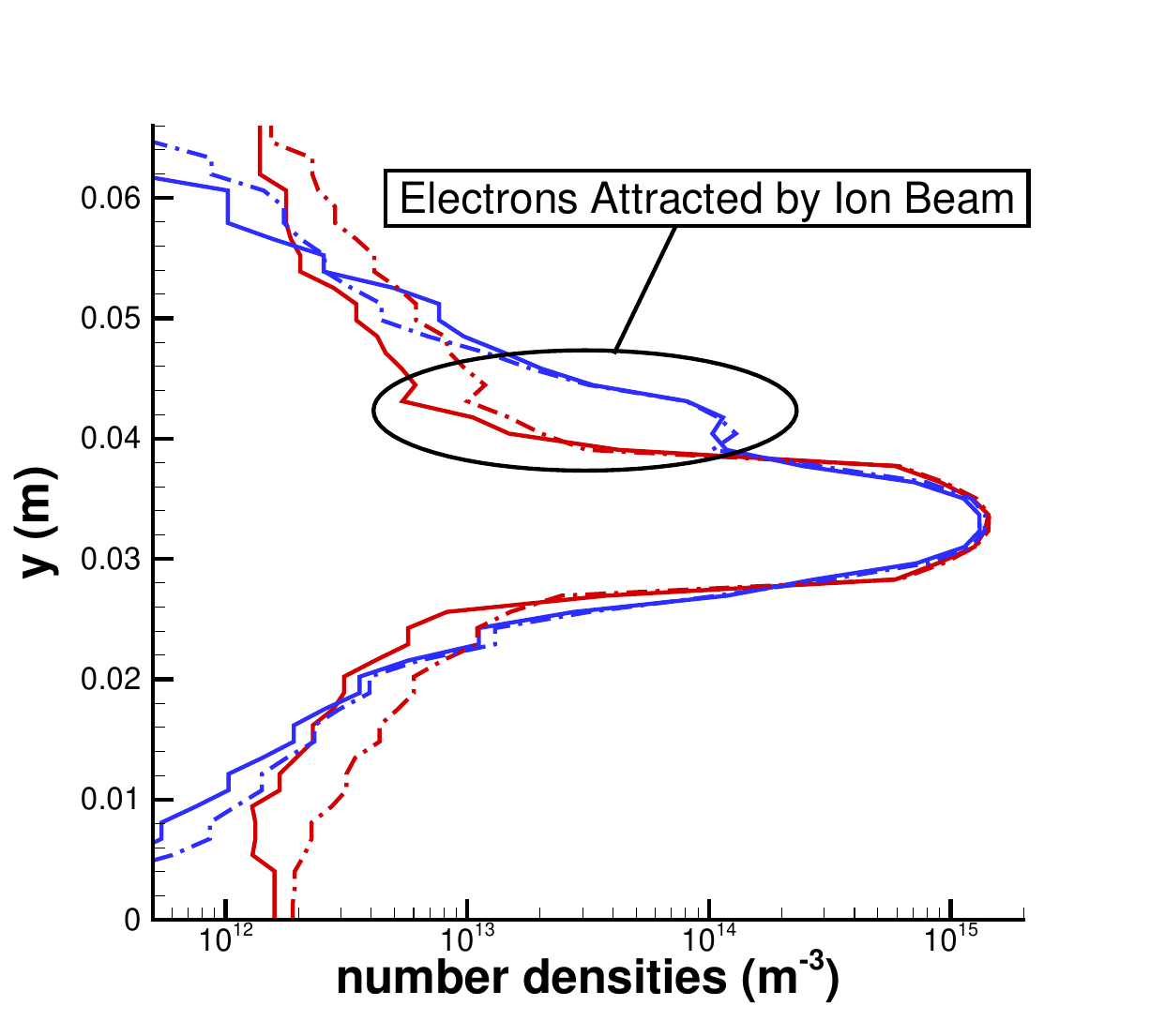}
		\includegraphics[width=0.45\linewidth]{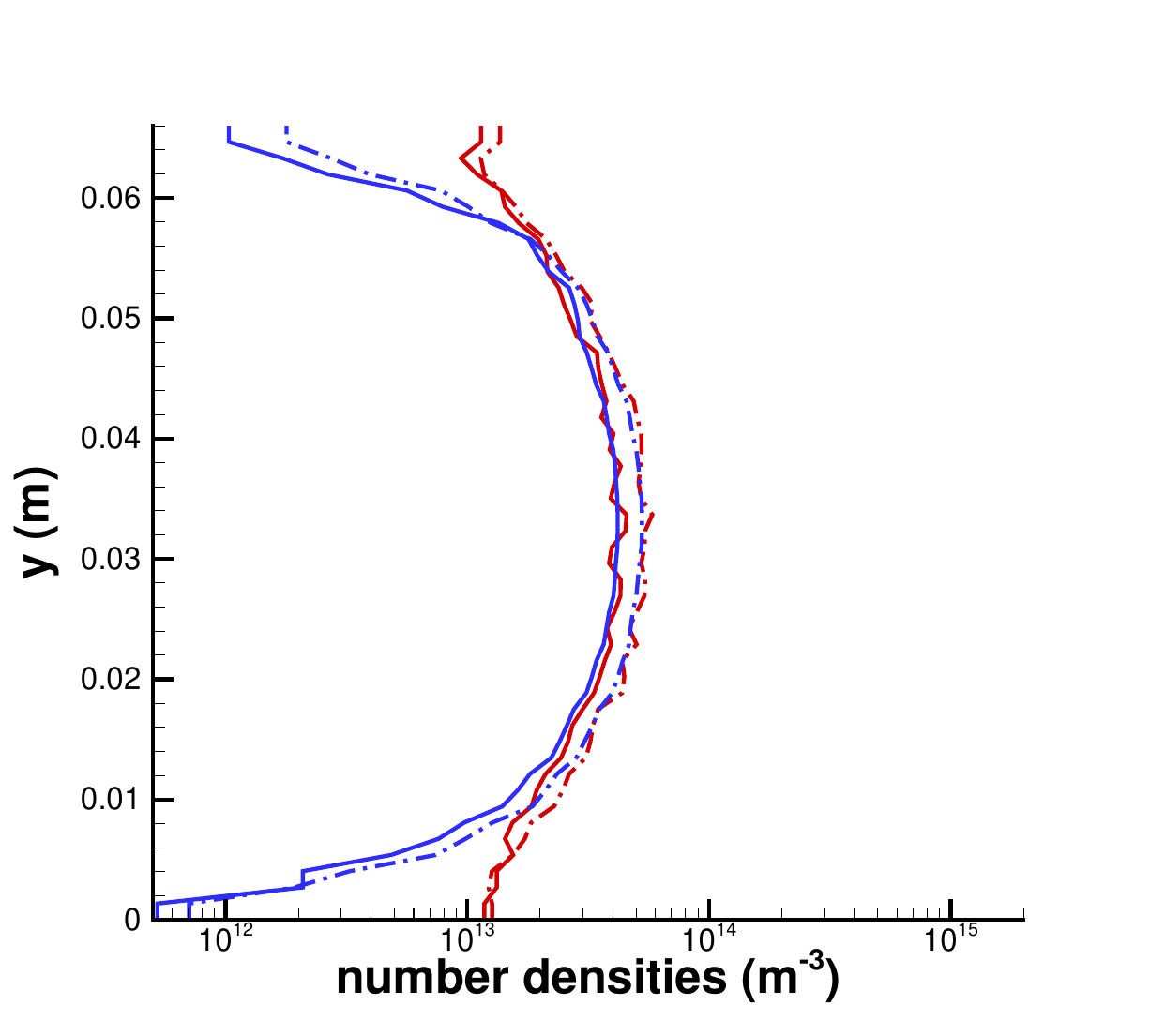}		
		\caption{Left: Number density profiles of ions (red) and electrons (blue) at $z = 0.01$~m (near-field) 
			for cases~1A (solid) and~1B (dash-dotted). Right: Number density profiles of ions (red) and electrons (blue) at $z = 0.1$~m (far-field) 
			for cases~1A (solid) and~1B (dash-dotted). }
		\label{Fig:10mm}		
		\vspace{0pt}
	\end{figure}
	
	To examine the effect of background pressure along the plume axis, 
	the plasma number density distributions of cases~1A and~1B are compared at two axial positions: 
	$z = 0.01$~m (near-field) and $z = 0.1$~m (far-field). 
	In the near-field region (left panel of Figure~\ref{Fig:10mm}), both cases exhibit a dense ion core, 
	but case~1B shows a higher ion number density in the beam-wing region due to enhanced CEX collisions at higher pressure. 
	Additionally, a noticeable increase in electron density is observed above the beam axis (positive $y$-direction), 
	indicating that electrons from the neutralizer are being attracted toward the positively charged ion beam.

	Further downstream at $z = 0.1$~m (right panel of Figure~\ref{Fig:10mm}), both ion and electron densities decrease as the plume expands into the far field. 
	However, case~1B maintains slightly higher overall plasma density, primarily because the increased background pressure produces a larger population of slower ions through CEX processes. These slower ions contribute to a denser and more sustained plume, highlighting the influence of pressure on the downstream plasma structure.

	\subsection{Beam characteristics}
	To characterize the beam properties, the beam current and divergence were evaluated at various axial positions \(z=z_0\) using the time-averaged solution. 
	The beam current through a plane at \(z=z_0\) is computed by integrating the axial current density, \(j_{\mathrm{beam}} = q\,n_i\,u_z\), as
	\begin{equation}
		I_B(z_0) = \int_{A(z=z_0)} j_{\mathrm{beam}}(x,y,z_0)\, dA,
		\label{eq:beam_current}
	\end{equation}
	where \(A(z_0)=\{(x,y):z=z_0\}\) denotes the sampling plane. 
	The local ion velocity components \((u_x, u_y, u_z)\) are obtained from the sampled data on this plane. 
	The beam divergence angle is then defined as
	\begin{equation}
		\lambda(z_0) = \cos^{-1}\!\left(
		\frac{\int_{A(z=z_0)} j_{\mathrm{beam}}
			\frac{u_z}{\sqrt{u_x^2 + u_y^2 + u_z^2}}\, dA}
		{I_B(z_0)}
		\right).
		\label{eq:beam_divergence}
	\end{equation}
	\begin{figure}[htbp]
		\centering
		\includegraphics[width=0.45\linewidth]{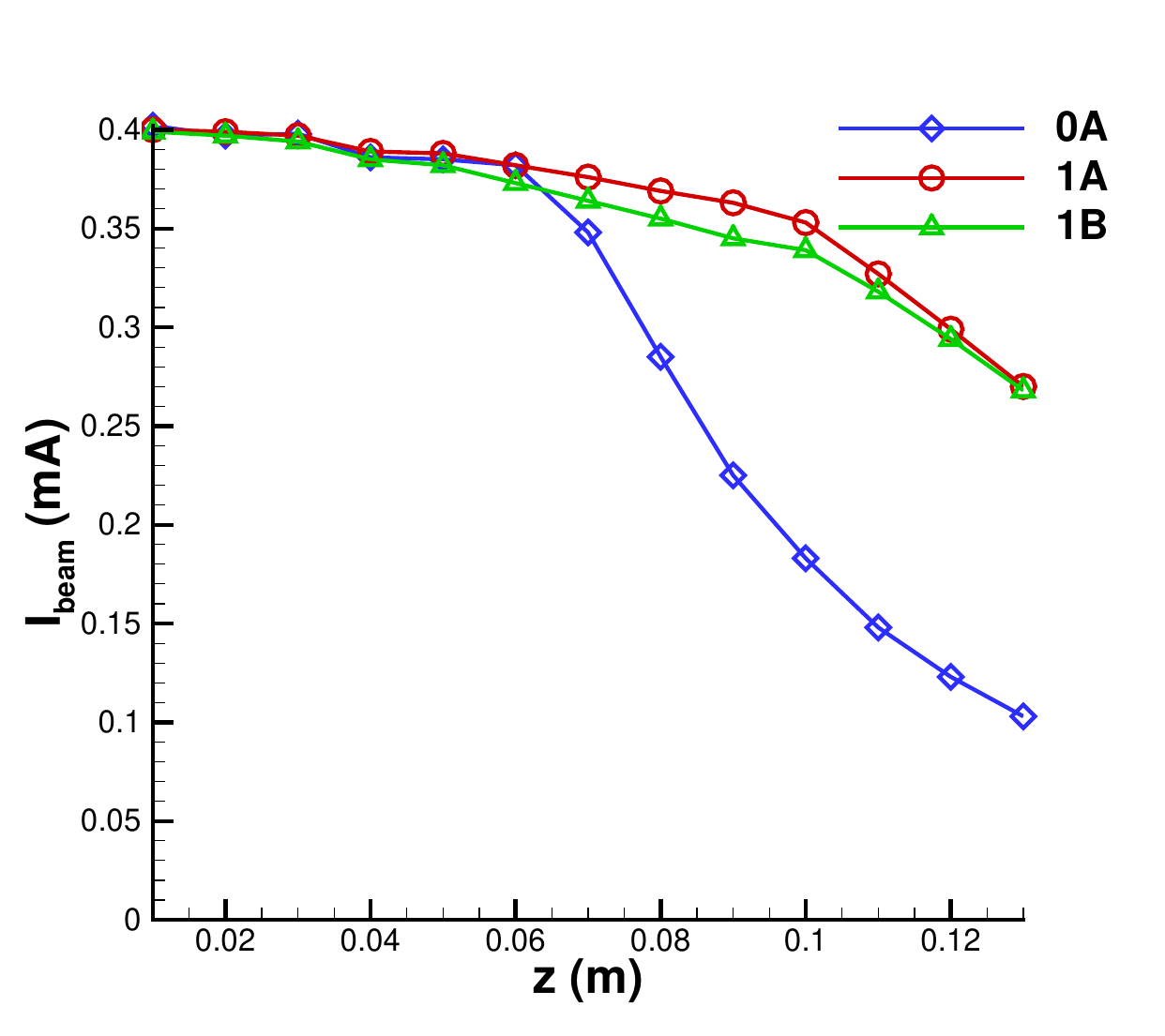}
		\includegraphics[width=0.45\linewidth]{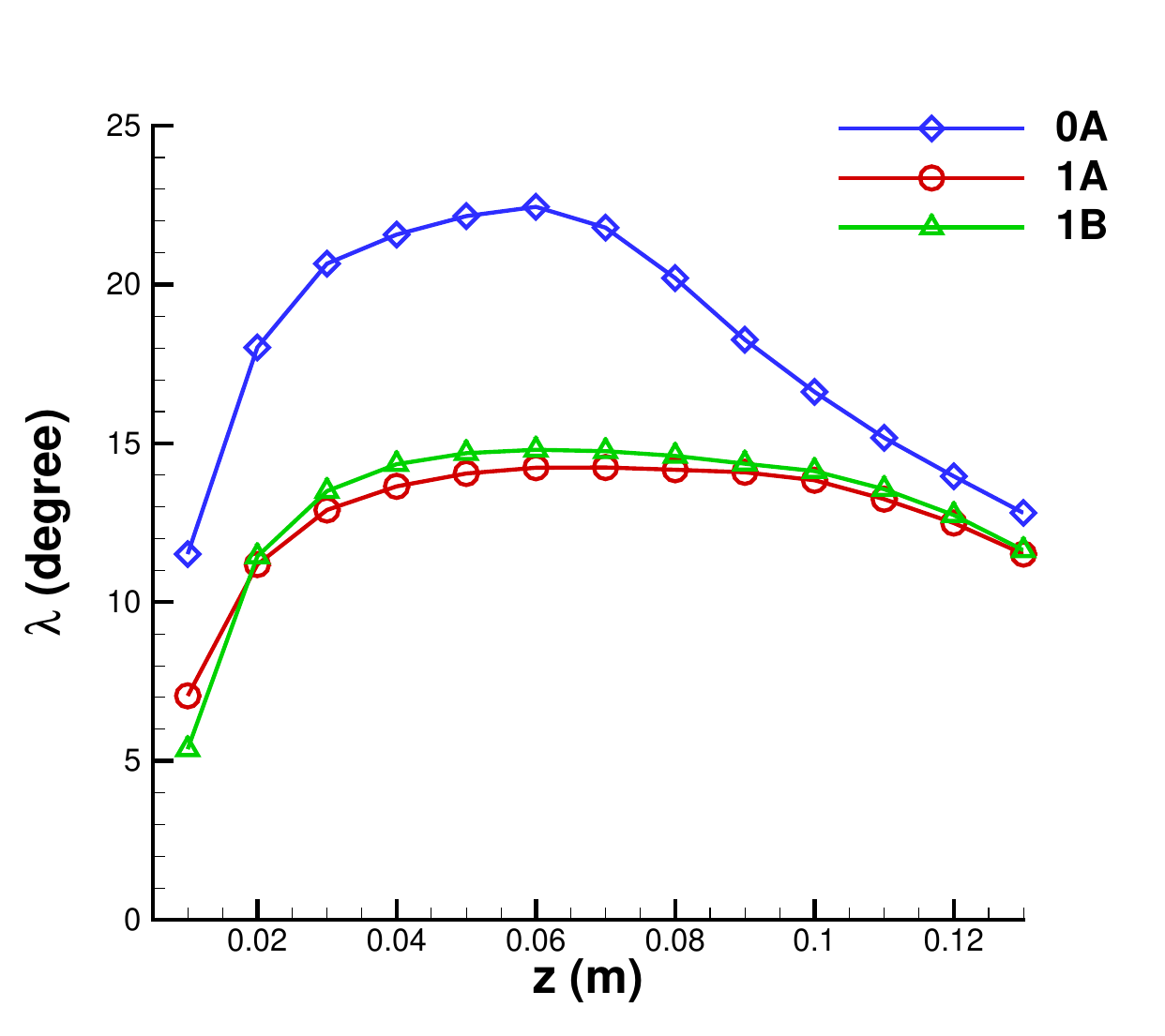}
		\caption{Beam current (left) and current-weighted divergence angle (right) versus axial position $z$, evaluated on planes from $z=0.01$ to $0.13~\mathrm{m}$ in $0.01~\mathrm{m}$ steps. For beam divergence from $z=0.07~\mathrm{m}$ (case~0A) and $z=0.09~\mathrm{m}$ (case~1A and 1B), values are shown for completeness but may be affected by wall loses. The apparent decrease in divergence angle after these positions results from ions with larger divergence angles being absorbed by the chamber walls, thereby reducing the contribution of off-axis ions to the computed beam divergence.}
		\label{Fig:beam}
	\end{figure}
	
	Figure~\ref{Fig:beam} shows the beam current and current-weighted divergence angle as functions of the axial position $z$, 
	evaluated on planes from $z = 0.01$ to $0.13$~m in increments of $0.01$~m. 
	Initially, the beam current remains close to the imposed value of $0.4~\mathrm{mA}$ for all cases, with only a slight decrease as the beam propagates downstream. 
	For case~0A, however, a sharp current drop occurs beyond $z = 0.06$~m, where the beam begins to impinge on the sidewalls due to its large divergence angle (bottom panel of Figure~\ref{Fig:beam}), resulting in significant ion losses. 
	The excessive divergence in case~0A originates from the strong potential gradient formed by the absence of inelastic collisions, 
	which in turn, prevents proper charge neutralization and produces an unrealistically large space-charge field. 
	Therefore, the beam spreading and wall losses observed in case~0A are not physically representative of actual gridded ion thruster operation. 
	In contrast, cases~1A and~1B maintain smaller divergence angles, keeping the beam more collimated and preserving the current farther downstream, up to about $z = 0.09$~m. 
	These results demonstrate that inelastic collisions and charge-exchange processes improve beam collimation by enhancing charge neutralization and suppressing space-charge divergence.

	When comparing cases~1A and~1B, the beam current of case~1B is slightly lower than that of case~1A, 
	which can be attributed to increased ion–neutral collisions at the higher background pressure. 
	Regarding beam divergence, case~1B initially exhibits a smaller divergence angle near the thruster exit ($z = 0.01$~m), 
	owing to the lower electric potential and reduced beam acceleration in the higher-pressure environment. 
	However, as the plume propagates, collisions with neutrals gradually increase the divergence, leading to a slightly broader beam in the downstream region.

	\subsection{Electron trajectories and trapping}
	\subsubsection{Electron trajectories of primary electrons}
	
	To better understand the electron dynamics inside the test chamber, 
	collisionless single-particle simulations were performed using the sampled electric potential field obtained from the PIC simulation case 1A. 
	Here, we define \textit{primary electrons} as those emitted from the neutralizer exit that have not undergone any inelastic collisions. 
	
\begin{figure}[htbp]
		\centering
		\includegraphics[width=0.45\linewidth, trim=0 50 0 20,clip]{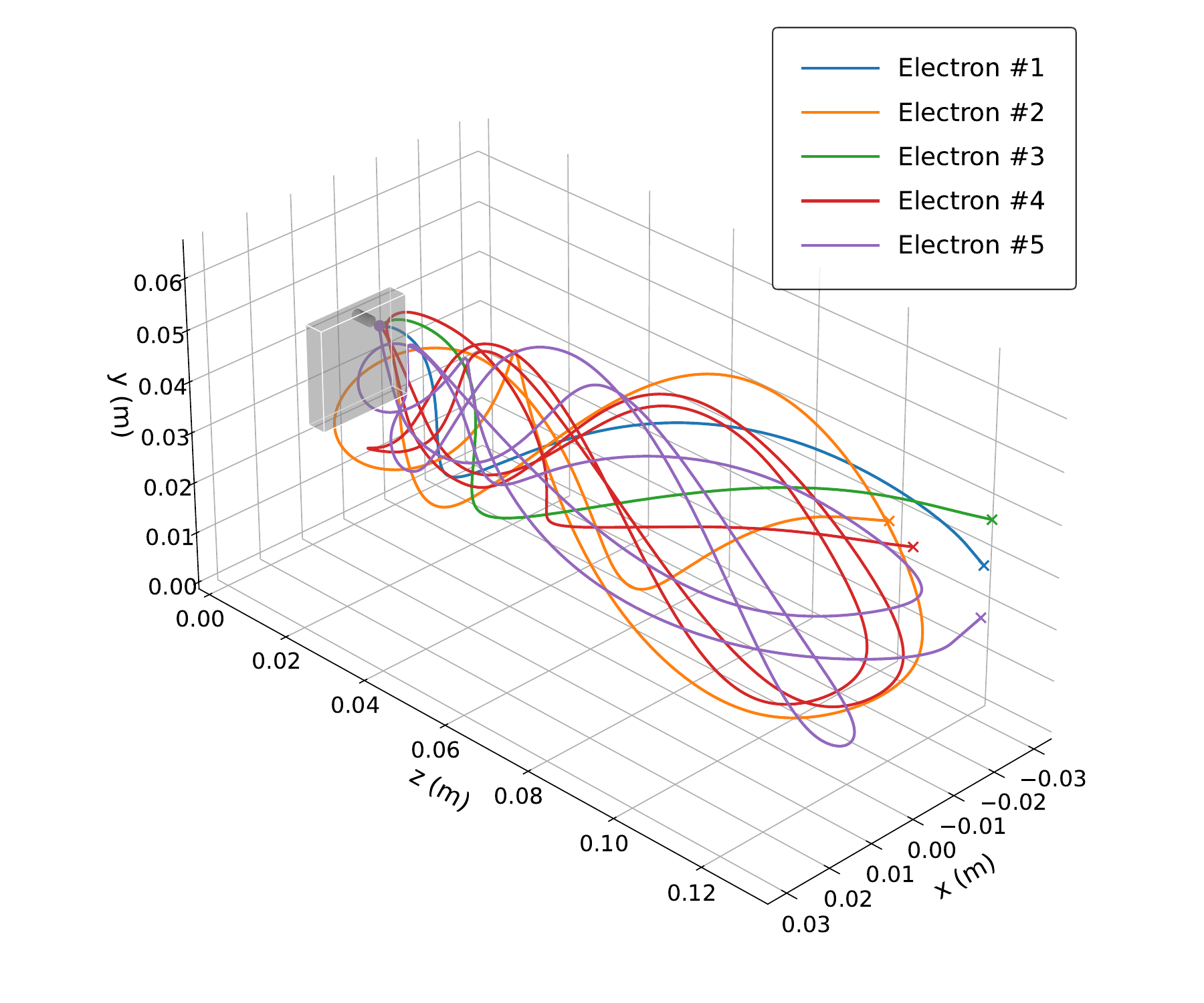}
		\includegraphics[width=0.45\linewidth]{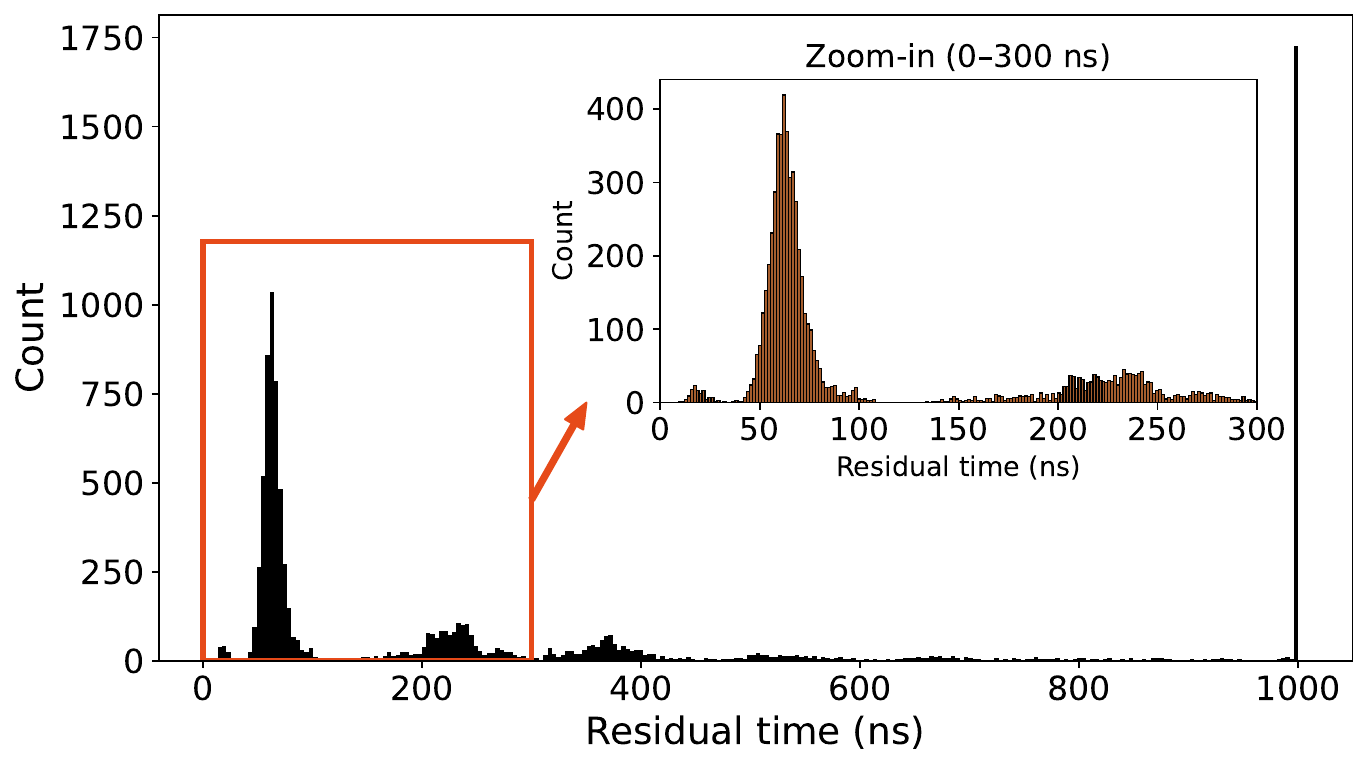}		
		\caption{Left: Representative trajectories of five primary electrons with an initial energy of $2~\mathrm{eV}$, showing both rapidly escaping and temporarily trapped behaviors. The gray structures indicate the plasma screen and neutralizer. Electrons~\#1--\#5 exited the domain at 60.47, 271.53, 66.74, 368.59, and 357.91~ns, respectively. Right: Residual-time distribution of 10,000 primary electrons with an initial energy of 2~eV. 
			The main plot shows that many electrons escape within the first 100~ns, while a small fraction remain longer and gradually escape over time. 
			The inset figure provides a zoomed-in view of the early-time distribution (0–200~ns), highlighting a strong peak corresponding to the rapid escape of electrons shortly after injection. Approximately 17.2\% of electrons remained within the domain at 1~$\mu$s, indicating temporary trapping in the electric potential field. Differences in the histogram counts between the main and inset plots arise from the use of different bin sizes and range definitions.}
		
		\label{Fig:primary_electron_traj}
	\end{figure}

	To obtain these trajectories, primary electrons were introduced near the neutralizer exit to reproduce the electron current emitted during operation. 
	A half-Maxwellian velocity distribution was applied in the axial $z$-direction, while full Maxwellian distributions were used in the $x$- and $y$-directions, corresponding to an average electron energy of $2~\mathrm{eV}$. In left panel of Figure~\ref{Fig:primary_electron_traj}, the trajectories of five representative primary electrons are shown. It can be seen that some electrons escaped rapidly, whereas others undergo multiple reflections within the potential well formed by the ion beam before eventually leaving the computational domain. 

	Right panel of Figure~\ref{Fig:primary_electron_traj} shows the residual-time distribution for 10,000 simulated primary electron trajectories. 
	It can be observed that many primary electrons escape the domain early in the simulation, while the remaining electrons continue to escape gradually over time. 
	When the maximum simulation time was set to $1~\mu\mathrm{s}$, approximately $17.2\%$ of the electrons remained, indicating that they were temporarily trapped within the local electric potential structure by the ion beam.These results demonstrate the coexistence of both quickly escaping and temporarily trapped electron populations, governed by the electric field topology near the thruster and neutralizer.

	\subsubsection{Post-inelastic collision electrons trajectories}
	\begin{figure}[htbp]
		\centering
		\includegraphics[width=0.7\linewidth, trim=80 150 0 150,clip]{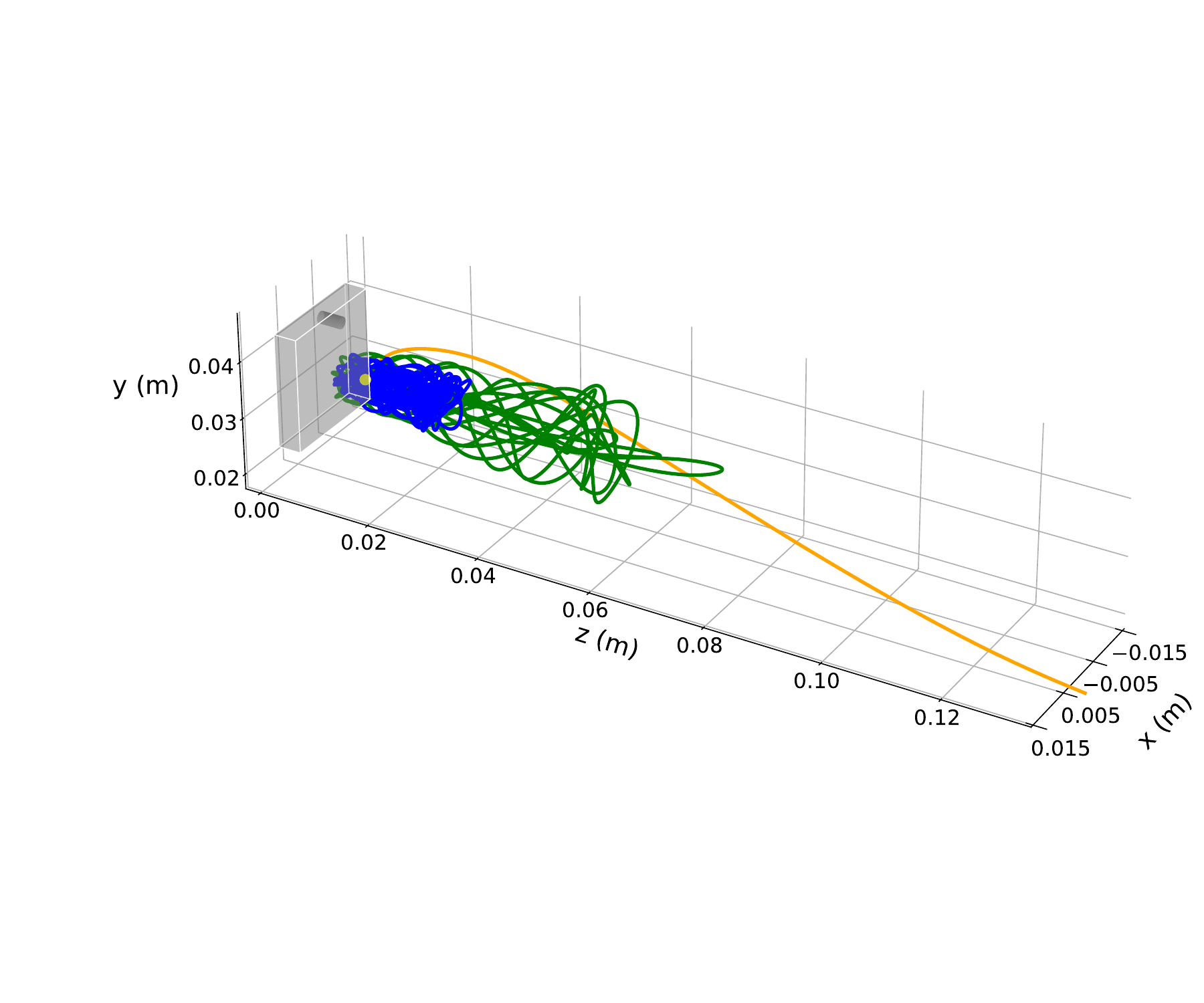}
		\caption{Representative trajectories of post-inelastic-collision electrons with different initial energies of 10 (blue), 20 (green), and 40~eV (orange). The gray structure represents the plasma screen and the neutralizer.}
		\label{Fig:trapoped_electron_traj}
	\end{figure}
	To better understand the motion of electrons generated after inelastic collisions, 
	we performed single-particle trajectory simulations of post-collision electrons using the steady state electric potential field of case 1A. 
	While the detailed statistics of the inelastic collision locations and post-collision energy distributions 
	are provided in Appendix~\ref{Appendix:inelastic_stats}, 
	it is important to note that the initial position of these electrons affects their potential energy within the chamber. 
	To minimize variables and isolate the effect of kinetic energy on trapping behavior, 
	we fixed the initial location of all post-collision electrons near the most probable collision site 
	(\(x, y, z \approx 0, 0.033, 0.01~\mathrm{m}\)), corresponding to a location on the centerline near the thruster exit.
	
	Using this setup, electrons were initialized with kinetic energies ranging from 5 to 80 eV, each following an isotropic velocity distribution consistent with a full Maxwellian corresponding to the specified kinetic energy.
	Three representative trajectories as shown in Fig.~\ref{Fig:trapoped_electron_traj}. 
	Low-energy electrons remain confined near the thruster exit due to stronger trapping within the local potential well, 
	while higher-energy electrons escape more easily. 
	The survival fraction, i.e., the fraction of electrons remaining in the simulation domain, after $1~\mu\mathrm{s}$ decreases with increasing energy, 
	indicating that higher-energy electrons contribute less to the steady-state neutralization cloud, 
	whereas low-energy electrons play a dominant role in maintaining local charge balance.
	
	\begin{table}[htbp]
		\centering
		\caption{Survived population of post-inelastic-collision electrons using 10,000 electron for each energy level after $1~\mu\mathrm{s}$ for different initial energies.}
		\label{Table:post_inelastic_survival}
		\begin{tabular}{cc}
			\hline
			Electron Energy (eV) & Survived Population (\%) \\
			\hline
			5  & 99.83 \\
			10 & 94.38 \\
			15 & 84.36 \\
			20 & 74.12 \\
			40 & 43.58 \\
			80 & 19.53 \\
			\hline
		\end{tabular}
	\end{table}
	As summarized in Table~\ref{Table:post_inelastic_survival}, 
	the survival fraction of electrons drops as the initial energy increases. 
	This trend shows that low-energy electrons generated from inelastic collisions 
	are much more likely to be trapped near the ion beam and thruster exit, 
	enhancing the probability of local charge accumulation and beam neutralization.

	\subsection{Sheath formation near vacuum chamber walls}
	\begin{figure}[htbp]
		\centering
		\includegraphics[width=0.7\linewidth]{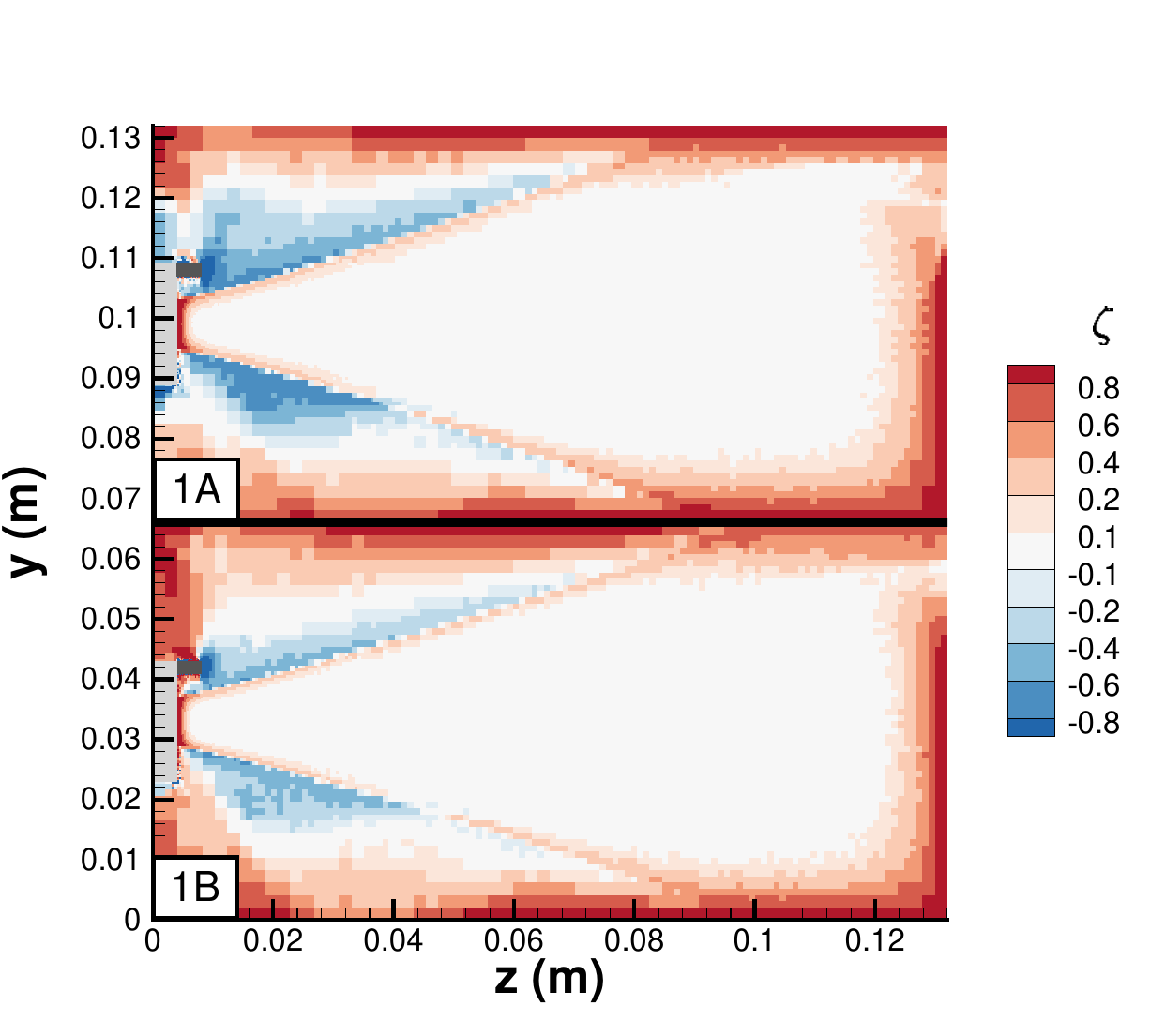}
		\caption{
			Relative charge imbalance ($\zeta$) contours in the $x=0$ plane for cases~1A (top) and~1B (bottom).
		}
		\label{Fig:1A1B-zeta}
	\end{figure}
	
	To examine sheath formation along the vacuum chamber walls, the local charge imbalance is used as a quantitative indicator of quasi-neutrality. 
	The relative charge imbalance is defined as
	\begin{equation}
		\zeta \;\equiv\; \frac{n_i - n_e}{n_i + n_e},
		\label{eq:zeta}
	\end{equation}
	where $\zeta\!\approx\!0$ (white), $\zeta>0$ (red), and $\zeta<0$ (blue) denotes quasi-neutral, ion-rich, and electron-rich regions, respectively. 
	
	Figure~\ref{Fig:1A1B-zeta} shows the sampled contours of $\zeta$ in the $x=0$ plane for cases~1A and~1B.  
	Electrons accumulate near the beam wings, attracted toward the positively charged ion beam. 
	A slight difference appears near the beam periphery, where case~1A exhibits a relatively higher electron density (i.e., more blue regions) compared to case~1B.  
	This behavior is consistent with the enhanced CEX ion production in the higher-pressure case~1B, which increases the population of slow ions near the beam edge and thereby improves local charge neutrality.
	Overall, both cases exhibit similar spatial distributions, indicating effective beam neutralization, with ion sheaths observed adjacent to the chamber side walls and beamdump ($z_{max}$).

	\begin{table}[htbp]
		\centering
		\caption{Plasma properties along the centerline at $(x, y) = (0.0, 0.033)~\mathrm{m}$ for cases~1A and~1B, corresponding to the sheath edge region ($\zeta = 0.05$) near the beamdump. The table compares the measured distance to the wall with the analytically evaluated sheath thicknesses based on the Child--Langmuir and Hutchinson models. $\lambda_{D}$, $s_\mathrm{data}$, $s_{CL}$, and $s_{H}$ are given in centimeters (cm).}
		
		\label{Table:sheath-BD}
		\begin{tabular}{cccccccc}
			\hline
			Case & $n_{i,s}$ ($\mathrm{m^{-3}}$) & $\phi_s$ (V) & $T_{e,s}$ (eV) & $\lambda_{D}$ & $s_\mathrm{data}$ & $s_{CL}$ & $s_{H}$ \\
			\hline
			1A & $2.99\times10^{13}$ & 19.135 & 4.882 & 0.301 & 1.484 & 0.847 & 1.131 \\
			1B & $3.56\times10^{13}$ & 16.186 & 3.852 & 0.244 & 1.374 & 0.726 & 1.000 \\
			\hline
		\end{tabular}
		\vspace{2pt}
		\begin{flushleft}
			\footnotesize\textit{Note.} $s_\mathrm{data}$ denotes the distance measured from the $z_\mathrm{max}$ sidewall along the $z$-direction toward the plasma region, corresponding to the sheath length from the PIC-MCC simulation.
		\end{flushleft}
	\end{table}

	To further investigate sheath formation, first we focus on the centerline, $(x, y) = (0.0, 0.033)~\mathrm{m}$, and examine the sheath formation at the beamdump ($z_{max}$). While Franklin~\cite{FranklinJPD2004} noted that it is difficult to specify an exact location for the sheath edge, he emphasized that authors should clearly state which point they refer to when discussing the plasma–sheath boundary.
	In this study, we define the sheath edge as the point where the charge imbalance parameter satisfies $\zeta > 0.05$ and continues to increase, marking the onset of quasi-neutrality breakdown. The corresponding plasma properties at the sheath edge along the centerline for cases~1A and~1B are summarized in Table~\ref{Table:sheath-BD}.
	
	To estimate the sheath thickness at this location, analytical models are used and compared. 
	According to the Child--Langmuir (CL) sheath model~\cite{HershkowitzPoP2005}, the sheath thickness normalized by the Debye length is expressed as
	\begin{equation}
		\frac{s_{\mathrm{CL}}}{\lambda_D}
		= \frac{0.79}{\sqrt{\alpha_i}}
		\left( \frac{\phi_s}{T_{e,s}} \right)^{3/4},
	\end{equation}
	where $\lambda_D$ is the Debye length, $\phi_s$ is the potential drop across the sheath, and $\alpha_i$ denotes the ratio of the ion density at the sheath edge to that in the quasi-neutral plasma ($\alpha_i = 0.61$ in this study). While the CL sheath corresponds to the regime where $\phi_s \gg T_{e,s}$, Hutchinson~\cite{Hutchinson2002} derived an analytical expression for the sheath thickness under the assumption that the electron density within the sheath is negligible and that the ion current density at the sheath edge equals the Bohm current. The normalized sheath thickness in Hutchinson model ($s_H$) is given by
	\begin{equation}
		\frac{s_{\mathrm{H}}}{\lambda_D}
		= 1.02
		\left[
		\left( \sqrt{\frac{\phi_s}{T_{e,s}}} - \frac{1}{\sqrt{2}} \right)^{1/2}
		\left( \sqrt{\frac{\phi_s}{T_{e,s}}} + \sqrt{2} \right)
		\right].
		\label{eq:hutchinson_sheath}
	\end{equation}
	Compared with the CL model, Hutchinson's formulation provides a more general description of the sheath, offering a smooth transition from the Debye sheath to the CL limit. It extends the applicability to regimes with relatively small potential drops ($\phi_s/T_e$), consistent with the present simulation conditions where the plasma-edge potential and electron temperature are comparable.

		\begin{figure}[htbp]
		\centering
		\includegraphics[width=0.7\linewidth]{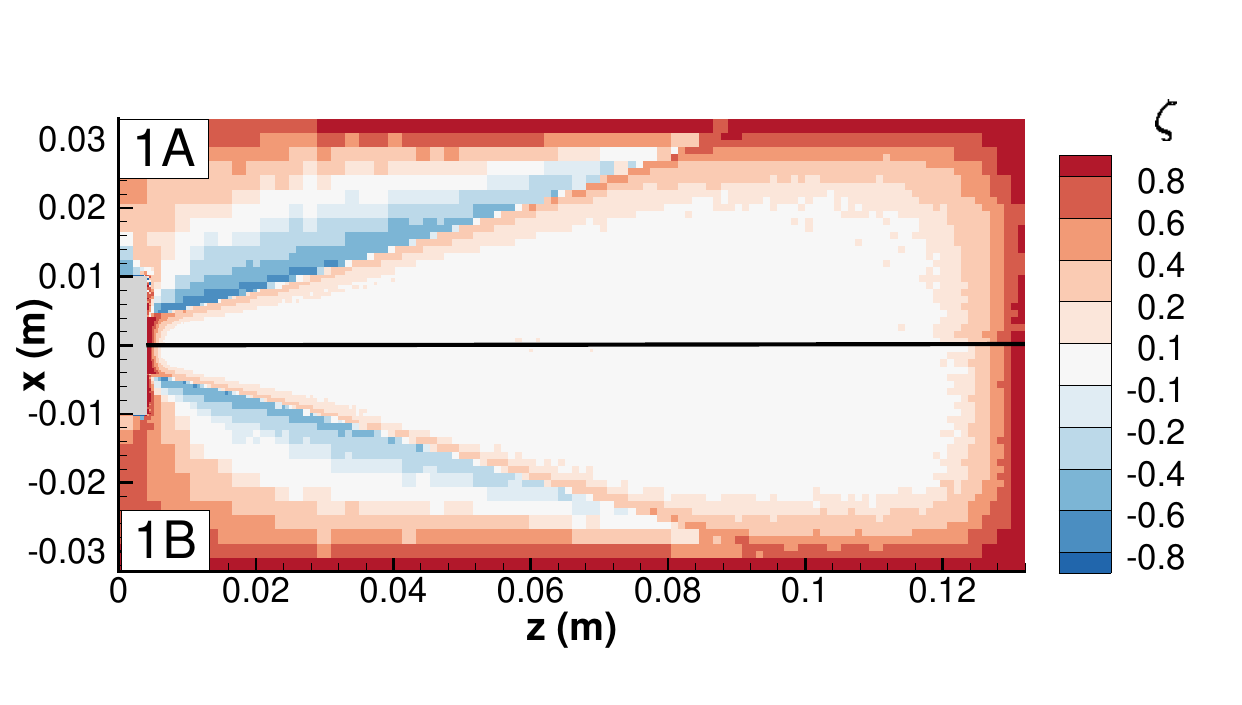}
		\caption{
			Relative charge imbalance ($\zeta$) contours in the $y=0.033$ plane for cases~1A (top) and~1B (bottom).
		}
		\label{Fig:1A1B-zeta2}
	\end{figure}

	Compared with the simulation, both analytical models underestimate the sheath length for cases~1A and~1B (Table~\ref{Table:sheath-BD}). In both cases we observe the ordering $s_{CL} < s_{H} < s_{\mathrm{data}}$: the CL model gives the smallest values, while Hutchinson’s formulation, which remains applicable for moderate $\phi_s/T_e$, reduces the discrepancy. Across theory and data, case~1B yields a shorter sheath than case~1A. This trend arises because the higher plasma density in case~1B leads to a smaller Debye length, and this effect dominates over the influence of $\phi_s/T_e$ under the present simulation conditions. 
	
	The overall underprediction of both analytical models can be attributed to the fact that a non-negligible fraction of electrons remains within the sheath region, leading to a weaker electric field and a more gradual potential gradient, thereby broadening the sheath compared with the idealized analytical models. Although not directly analogous, previous studies~\cite{HatamiSR2022, GyergyekPoP2005, DemidovPRL2005} have similarly reported that the presence of electrons in the sheath region can result in increased sheath thickness.

	To analyze sheath formation near the sidewall, we adopt a similar methodology to the centerline analysis.  Figure~\ref{Fig:1A1B-zeta2} shows the sampled contours of $\zeta$ in the $y=0.033~\mathrm{m}$ plane for cases~1A and~1B.  The sampling location is fixed at $(y, z) = (0.033, 0.025)~\mathrm{m}$, which corresponds to the sheath edge region near the sidewall. At this location, plasma parameters are sampled along the $x$-direction, from the wall into the bulk plasma, and the sheath length is estimated from the local plasma profiles. Table~\ref{Table:sheath-side} summarizes the key plasma parameters and compares the measured sheath length with analytical predictions from the CL and Hutchinson models.
	\begin{table}[htbp]
		\centering
		\caption{Plasma properties along the line at $(y, z) = (0.033, 0.025)~\mathrm{m}$ for cases~1A and~1B, corresponding to the sheath edge region ($\zeta = 0.05$) near the sidewall. The table compares the measured distance to the wall with the analytically evaluated sheath thicknesses based on the Child--Langmuir and Hutchinson models. $\lambda_{D}$, $s_\mathrm{data}$, $s_{CL}$, and $s_{H}$ are given in centimeters (cm).}
		\label{Table:sheath-side}
		\begin{tabular}{cccccccc}
			\hline
			Case & $n_{i,s}$ ($\mathrm{m^{-3}}$) & $\phi_s$ (V) & $T_{e,s}$ (eV) & $\lambda_{D}$ & $s_\mathrm{data}$ & $s_{CL}$ & $s_{H}$ \\
			\hline
			1A & $3.06\times10^{12}$ & 7.919 & 3.372 & 0.780 & 1.027 & 1.497 & 2.131 \\
			1B & $6.53\times10^{12}$ & 9.501 & 2.982 & 0.502 & 1.236 & 1.212 & 1.702 \\
			\hline
		\end{tabular}
		\vspace{2pt}
		\begin{flushleft}
			\footnotesize\textit{Note.} $s_\mathrm{data}$ denotes the distance measured from the $x_\mathrm{max}$ sidewall along the $x$-direction toward the plasma region, corresponding to the sheath length.
		\end{flushleft}
	\end{table}

\begin{table*}[htbp]
	\caption{Percentage of ion and electron currents collected at each boundary and 
		their corresponding mean energies for cases~1A and~1B after reaching steady state. 
		All current values are expressed in percent of the \textit{total outflow} ion and 
		electron currents, $I_{i,\mathrm{out}}$ and $I_{e,\mathrm{out}}$, respectively. 
		Mean energies, $\langle E_e \rangle$, $\langle E_{i,\mathrm{fast}} \rangle$, 
		and $\langle E_{i,\mathrm{slow}} \rangle$, are given in electronvolts (eV). 
		The total outflow currents represent the net currents exiting the computational domain.}
	\label{Table:current_path}
	\begin{ruledtabular}
		\begin{tabular}{lcccccc cccccc}
			\multirow{2}{*}{Boundary} 
			& \multicolumn{6}{c}{Case~1A} 
			& \multicolumn{6}{c}{Case~1B} \\
			\cline{2-7}\cline{8-13}
			& $I_e$ & $\langle E_e \rangle$ 
			& $I_{i,\mathrm{fast}}$ & $\langle E_{i,\mathrm{fast}} \rangle$
			& $I_{i,\mathrm{slow}}$ & $\langle E_{i,\mathrm{slow}} \rangle$
			& $I_e$ & $\langle E_e \rangle$ 
			& $I_{i,\mathrm{fast}}$ & $\langle E_{i,\mathrm{fast}} \rangle$
			& $I_{i,\mathrm{slow}}$ & $\langle E_{i,\mathrm{slow}} \rangle$ \\
			\hline
			$z_\mathrm{max}$  & 60.92 & 6.627 & 60.42 & 797.7 & 4.031 & 11.79
			& 60.17 & 6.273 & 54.94 & 784.0 & 6.505 & 9.982 \\
			$z_\mathrm{min}$  & 6.119 & 2.896 & 0.033 & 56.23 & 0.205 & 32.33
			& 2.285 & 3.233 & 0.053 & 45.41 & 0.378 & 23.83 \\
			$y_\mathrm{max}$  & 4.725 & 5.725 & 6.199 & 666.9 & 2.814 & 24.10
			& 6.083 & 5.939 & 5.518 & 523.3 & 4.228 & 18.85 \\
			$y_\mathrm{min}$  & 3.908 & 4.675 & 6.123 & 665.6 & 2.755 & 23.89
			& 5.018 & 6.522 & 5.292 & 517.9 & 4.291 & 18.59 \\
			$x_\mathrm{max}$  & 2.770 & 4.540 & 5.971 & 663.5 & 2.610 & 24.29
			& 3.107 & 5.522 & 4.960 & 516.9 & 4.415 & 18.85 \\
			$x_\mathrm{min}$  & 2.770 & 4.540 & 5.971 & 663.5 & 2.610 & 24.29
			& 3.107 & 5.522 & 4.960 & 516.9 & 4.415 & 18.85 \\
			Plasma screen     & 0.722 & 2.575 & 0.128 & 60.05 & 0.120 & 34.22
			& 0.346 & 3.019 & 0.171 & 52.73 & 0.212 & 26.23 \\
			Neutralizer       & 18.06 & 2.076 & 0.006 & 144.7 & 0.003 & 32.12
			& 19.88 & 2.097 & 0.012 & 146.6 & 0.011 & 24.14 \\
			\hline
			\textbf{Total}    & 100.0 &       & 84.85 &       & 15.15 &
			& 100.0 &       & 75.55 &      & 24.45 &       \\
		\end{tabular}
	\end{ruledtabular}
	
	\vspace{1ex}
	\begin{flushleft}
		\footnotesize\textit{Notes.}
		(1) $I_{i,\mathrm{fast}}$ includes beam and MEX ions, whereas 
		$I_{i,\mathrm{slow}}$ represents ions newly generated by CEX or ionization. \\
		(2) The total current collected at $x_\mathrm{max}$ represents contributions from 
		both $x_\mathrm{max}$ and $x_\mathrm{min}$ due to the symmetry about $x=0$; hence, it is 
		equally divided between the two boundaries, which share the same mean energy.
	\end{flushleft}
\end{table*}

	In both cases, the Hutchinson sheath prediction again remains larger than the CL model, as expected due to its inclusion of moderate $\phi_s/T_e$ effects. However, in contrast to the centerline analysis, the measured sheath length $s_\mathrm{data}$ is smaller than both analytical predictions. This lower prediction suggests that the simulation domain was insufficient to fully resolve the sheath structure, likely due to interference between the expanding ion beam and the sidewall boundary, resulting in mutual perturbation between the sheath and beam dynamics. Consequently, the physical sheath may have been truncated, leading to an underestimation of $s_\mathrm{data}$.

	These observations indicate the need for a larger simulation domain or chamber size in the lateral direction, to properly capture sheath formation without edge-induced artifacts. Although geometric scaling was applied based on thruster dimensions, it appears insufficient for accurately replicating sheath behavior, since sheath characteristics are not governed purely by physical dimensions, but are more strongly influenced by plasma properties. This highlights that dimensional scaling must be coupled with plasma-specific constraints, particularly sheath physics, to ensure fidelity when extrapolating to reduced or altered geometries, and it naturally motivates future studies toward full-scale simulations.

	\subsection{Current flow paths}
	
	Table~\ref{Table:current_path} summarizes the ion and electron current distributions collected at each boundary for cases~1A and~1B, based on all particles crossing each boundary after reaching steady state. While the injected ion and electron currents are 0.40~mA and 1.19~mA, respectively, the total outflow currents increase slightly to 0.41~mA and 1.20~mA for both case 1A and 1B. This small increase results from additional charged-particle generation through ionization processes within the domain. 
	For both cases, the downstream boundary ($z_\mathrm{max}$) dominates the total current flow, carrying over 60\% of the outgoing electrons and ~64.5\% of ions for case 1A  and ~61.5\% for case 1B.  Still, the current flows to sidewalls ($x_\mathrm{max/min}$, $y_\mathrm{max/min}$) is not negligible for both cases.

	In the present simulation, a considerable fraction of the emitted electron current is collected at the neutralizer surface, approximately 18 to 20\% of the total outflow current.
	A constant electron injection current is prescribed at the cathode boundary, following the simplified treatment commonly adopted in previous works \cite{NishiiJPP2023, NishiiPSST2023, GuaitaPSST2025}.
	This assumption is necessary because the detailed cathode–plume coupling, including the internal plasma dynamics and current modulation within the neutralizer, is highly complex and cannot be accurately resolved within the present model framework.
	The cathode–plume interaction itself is inherently nonlinear and time dependent; hence, the actual electron emission from the neutralizer involves complex internal dynamics, including temperature-driven oscillations and potential instabilities \cite{GoebelJAP2021}.
	As a result, a portion of the injected electrons is electrostatically redirected back toward the neutralizer, producing a finite return current.
	This return current is therefore thought to result from the constant-injection assumption rather than from the detailed neutralizer physics.
	
	At higher pressure (case~1B), both the fast and slow ion mean energies decrease across all boundaries.  
	The mean energy of fast ions at the plume exit drops from approximately $798~\mathrm{eV}$ to $784~\mathrm{eV}$, while that of slow ions decreases from $\sim12~\mathrm{eV}$ to $\sim10~\mathrm{eV}$.  
	This overall reduction results from more frequent ion--neutral interactions, which dissipate ion energy to background neutrals through MEX collisions.  
	The current partitioning further supports this collisional behavior: the total fraction of slow ions increases from 15.15\% in case~1A to 24.45\% in case~1B, indicating enhanced production of low-energy ions by CEX and ionization processes.  
	Correspondingly, the sidewall currents ($x_\mathrm{max/min}$ and $y_\mathrm{max/min}$) show a slight increase (about 3\% for both ions and electrons), and this trend is consistent with the expectation that higher pressure conditions promote limited beam scattering through cumulative CEX events, which marginally reduce ion directionality and redirect ions toward the sidewalls~\cite{NishiiJPP2023}.
	As a result, sidewall collection becomes slightly more pronounced, although the overall current transport remains predominantly axial, as also reflected in the beam current profiles of Figure~\ref{Fig:beam}, where case~1B exhibits faster current attenuation due to increased beam spreading and more ion losses to the charge-absorbing sidewalls.
	
	Even with the thruster exit and plasma screen boundaries grounded, the electron backstreaming current remains effectively suppressed, with only about 0.3 to 0.7\% of the total electron current collected at the plasma screen.  
	This indicates that the overall potential distribution near the thruster exit successfully confines electrons and prevents significant backflow into the discharge region.  
	If a negative bias voltage to the thruster exit had been applied, the backstreaming current would have been expected to further decrease, although slow ions will be increasingly attracted toward the negatively biased surface.

	\section{Conclusions}
	
	This study presented three–dimensional, fully kinetic PIC–MCC simulations informed by a DSMC neutral background to investigate facility effects on the plume of a reduced–scale gridded ion thruster in a ground–test environment. We showed that including electron–neutral inelastic processes is necessary to obtain a physically consistent, neutralized beam.  In fact, without inelastic cooling, unrealistically large potential gradients formed near the thruster exit, producing excessive lateral electric fields, centerline ion depletion, and premature wall losses in case 0A.  With inelastic processes enabled (cases~1A and~1B), the potential field flattens, ion and electron populations remain collimated, and the beam current is preserved farther downstream.
	
	Varying the background pressure from low to high levels primarily increases collisionality, leading to enhanced ion–neutral interactions that cool and slightly broaden the plume. The higher-pressure case exhibits a consistent but modest reduction in both fast- and slow-ion energies and a redistribution of the total current toward a larger fraction of slow ions, reflecting the enhanced role of CEX collisions. In addition, the potential field becomes more flattened at higher pressure with more CEX collisions, which reduce the overall electric-field strength near the thruster exit. 
	
	Electron trajectory analyses clarify the neutralization physics underlying these trends. Primary electrons exhibit a mix of rapid escape and temporary trapping in the beam–induced potential well, while post–inelastic electrons are preferentially formed at low energy, sustaining the neutralization cloud. Sheath diagnostics further reveal that, at the beam dump, both the Child–Langmuir and Hutchinson models underpredict the simulated sheath length ($s_{CL}<s_H<s_\mathrm{data}$), consistent with a non-negligible electron population persisting inside the sheath. Near the sidewall, however, the measured sheath is smaller than both analytical predictions, likely reflecting truncation by beam–sheath interference in the laterally compact domain. These findings underscore that chamber–scale geometry must be considered together with plasma kinetics to capture wall coupling and sheath structure. Current-path analyses further show that increasing background pressure enhances charge return to the sidewall while maintaining effective beam neutralization near the neutralizer. The overall current distribution indicates that electron loss to the sidewall remains limited, with most return current passing through  the beam core and dump, reinforcing the observed sheath asymmetry and wall coupling behavior. 
	
	Overall, the results highlight a coupled physical picture: inelastic electron cooling enables physically realistic beam neutralization; increased pressure enhances CEX, lowers ion and electron energies, slightly broadens the beam divergence, and increases sidewall losses leading to current attenuation. The sheath behavior is found to be sensitive to both residual electron populations and beam–wall proximity. Future work will focus first on improving the model fidelity through refined wall boundary conditions including electrical boundary condition and incorporation of secondary electron emission effects. Subsequently, full-scale simulations will be conducted to compare directly with experimental measurements and to systematically assess the influence of dimensional scaling on plume dynamics and facility-induced phenomena.

	\section*{Acknowledgments}
	This work was supported by NASA through the Joint Advanced Propulsion Institute, a NASA Space Technology Research Institute, Grant Number~80NSSC21K1118. 
	The authors would like to thank Dr.~Foster at the University of Michigan and his students, Dr.~Topham and Sophia~Bergmann, for their valuable discussions and collaborative efforts related to this work. 
	The authors also acknowledge Brett~Bode at the National Center for Supercomputing Applications (NCSA) for his technical assistance and support with the Delta computing resources. 
	This research used Delta at NCSA through allocation~PHY220158.

	\appendix
	\renewcommand{\thefigure}{A\arabic{figure}}
	\setcounter{figure}{0}
	\renewcommand{\thetable}{A\arabic{table}}
	\setcounter{table}{0}
	\section{Inelastic collision data}
	\label{Appendix:inelastic_stats}
	
	\begin{figure}[htbp]
		\centering
		\includegraphics[width=0.48\linewidth]{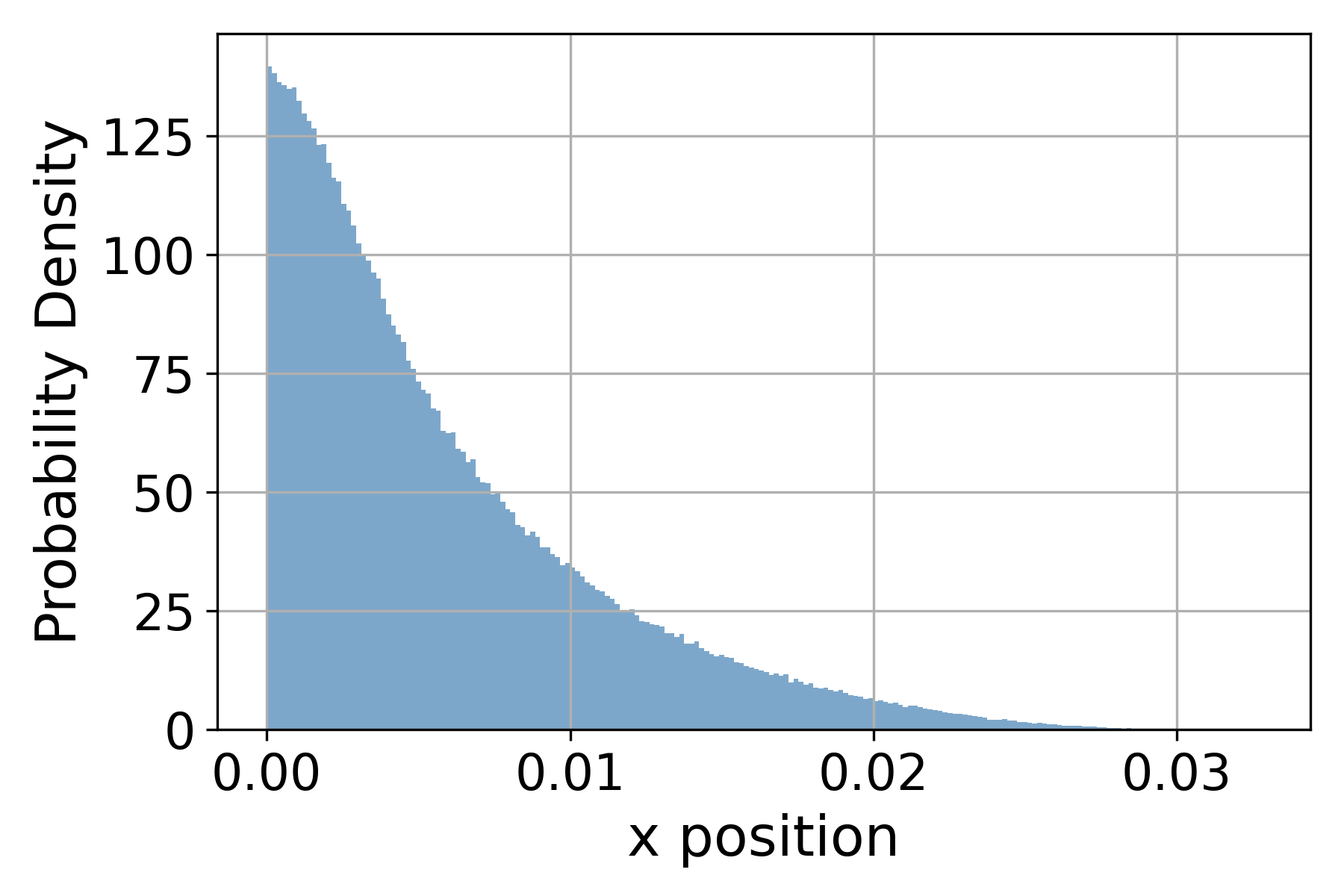}
		\includegraphics[width=0.48\linewidth]{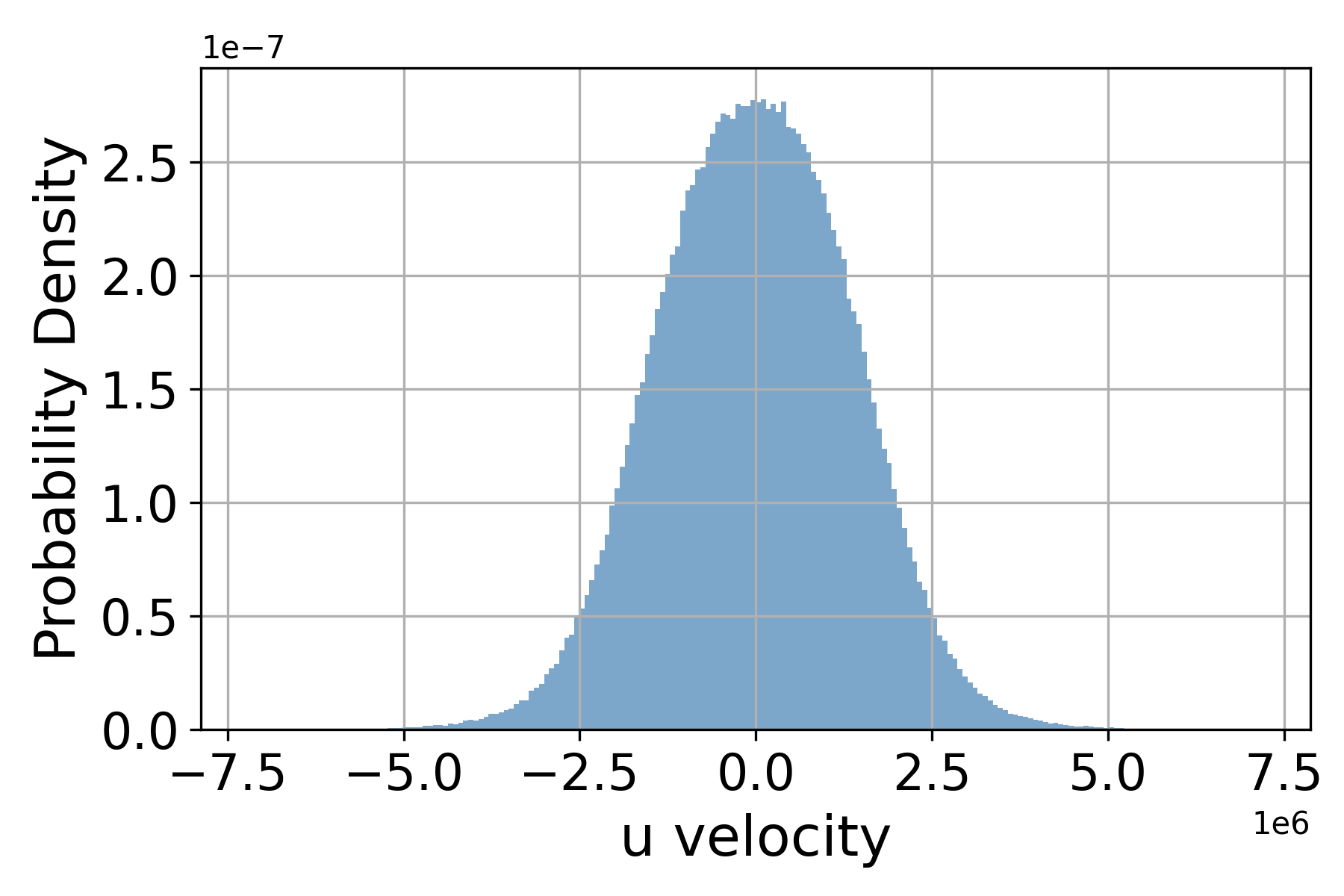}	
		\includegraphics[width=0.48\linewidth]{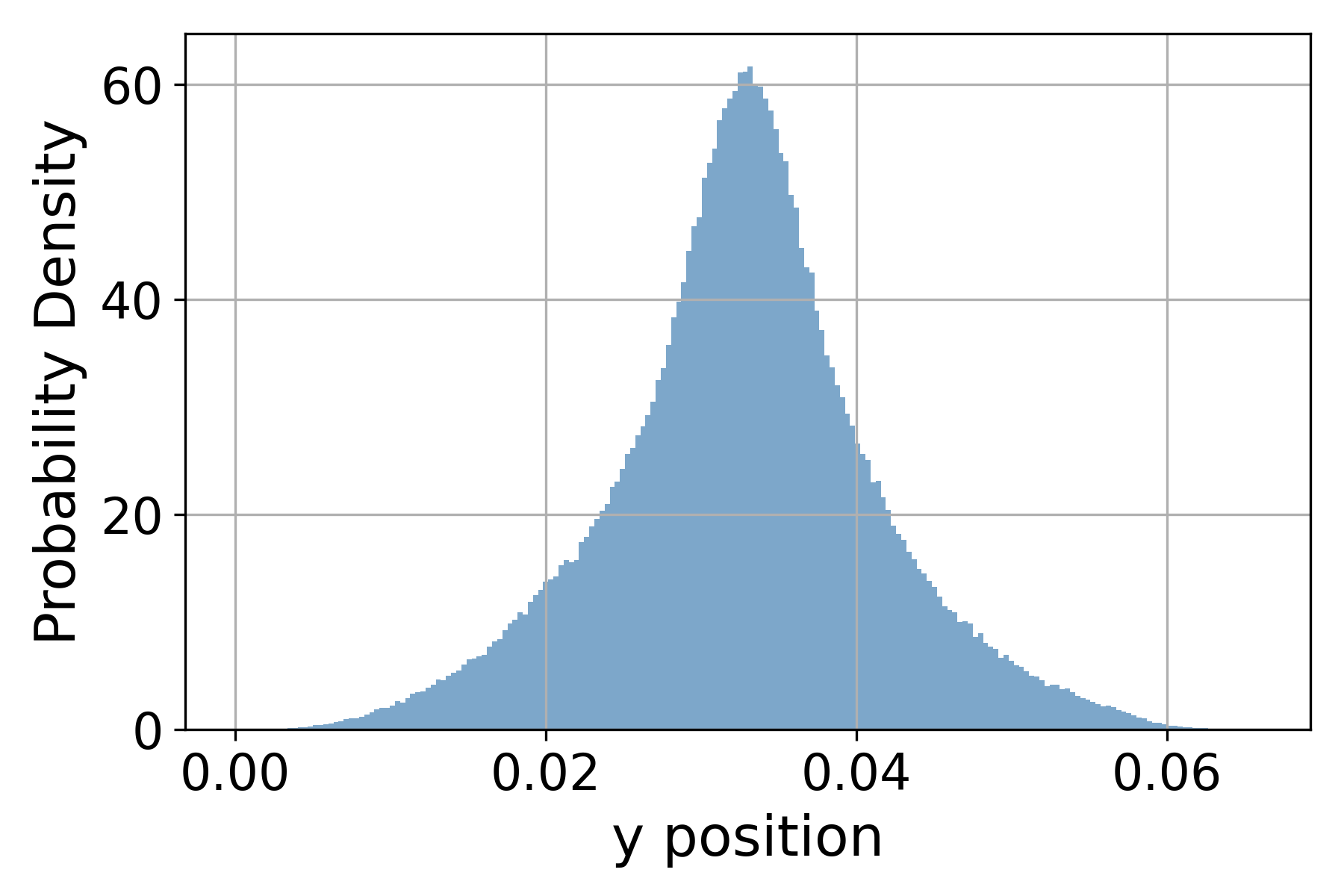}
		\includegraphics[width=0.48\linewidth]{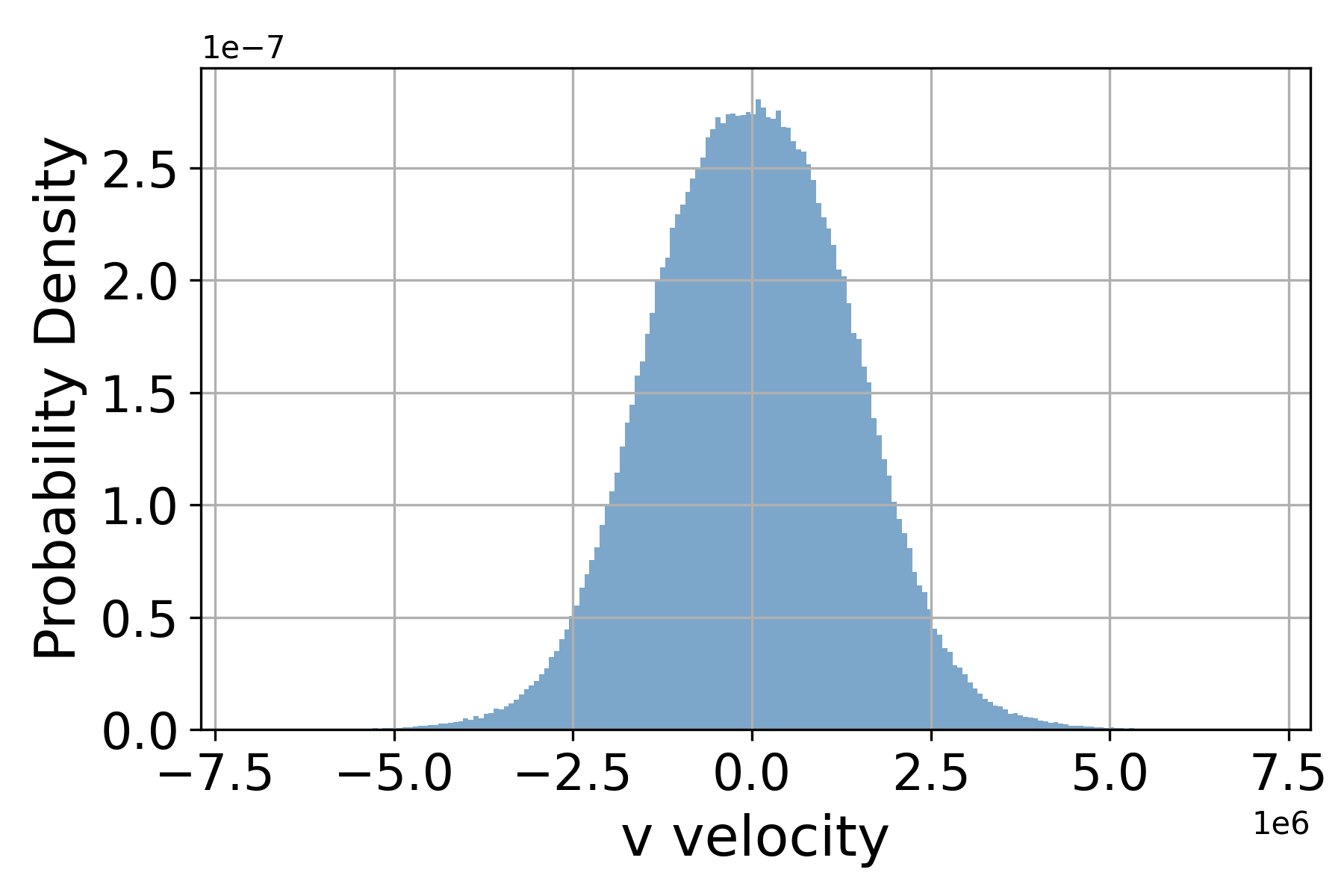}
		\includegraphics[width=0.48\linewidth]{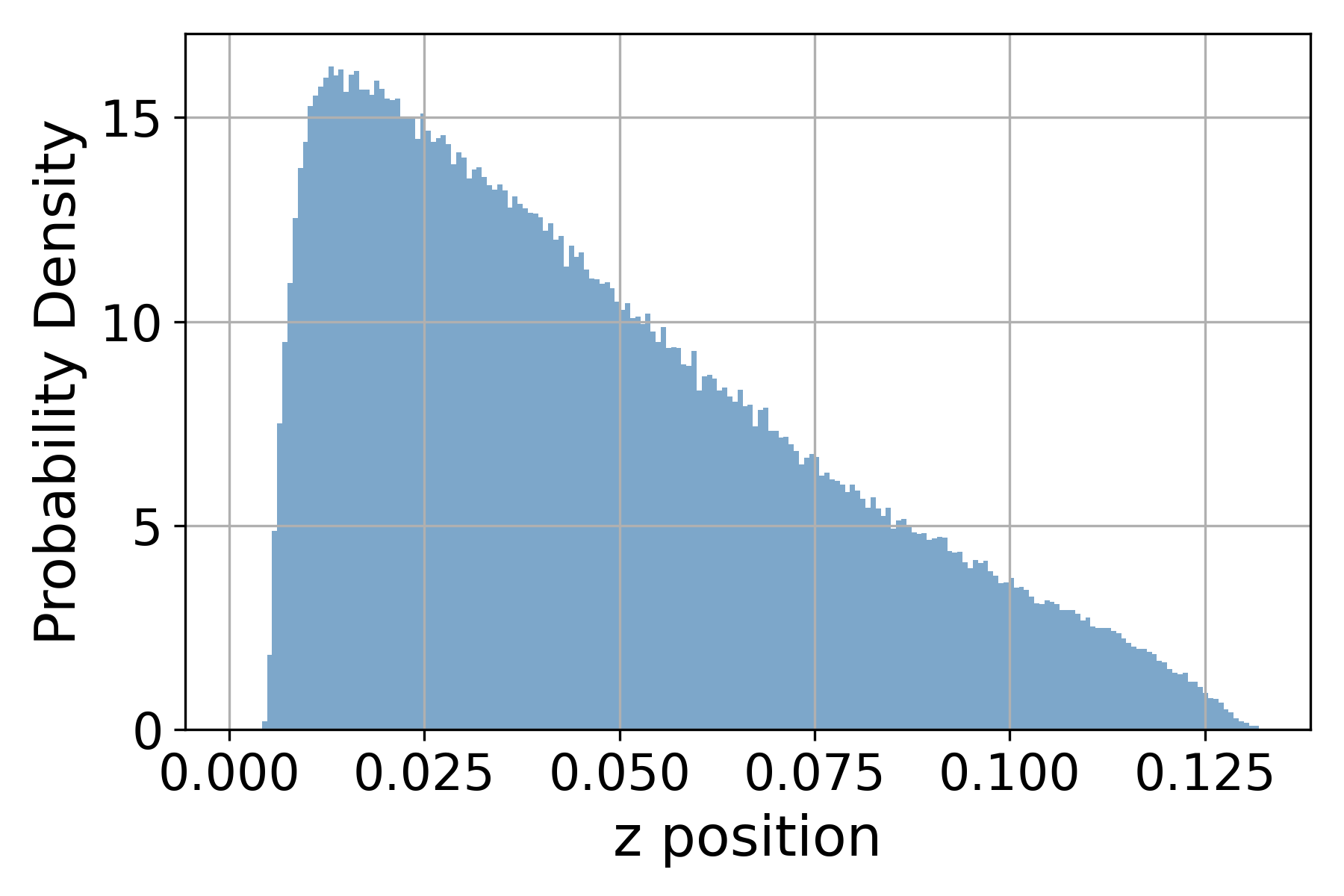}	
		\includegraphics[width=0.48\linewidth]{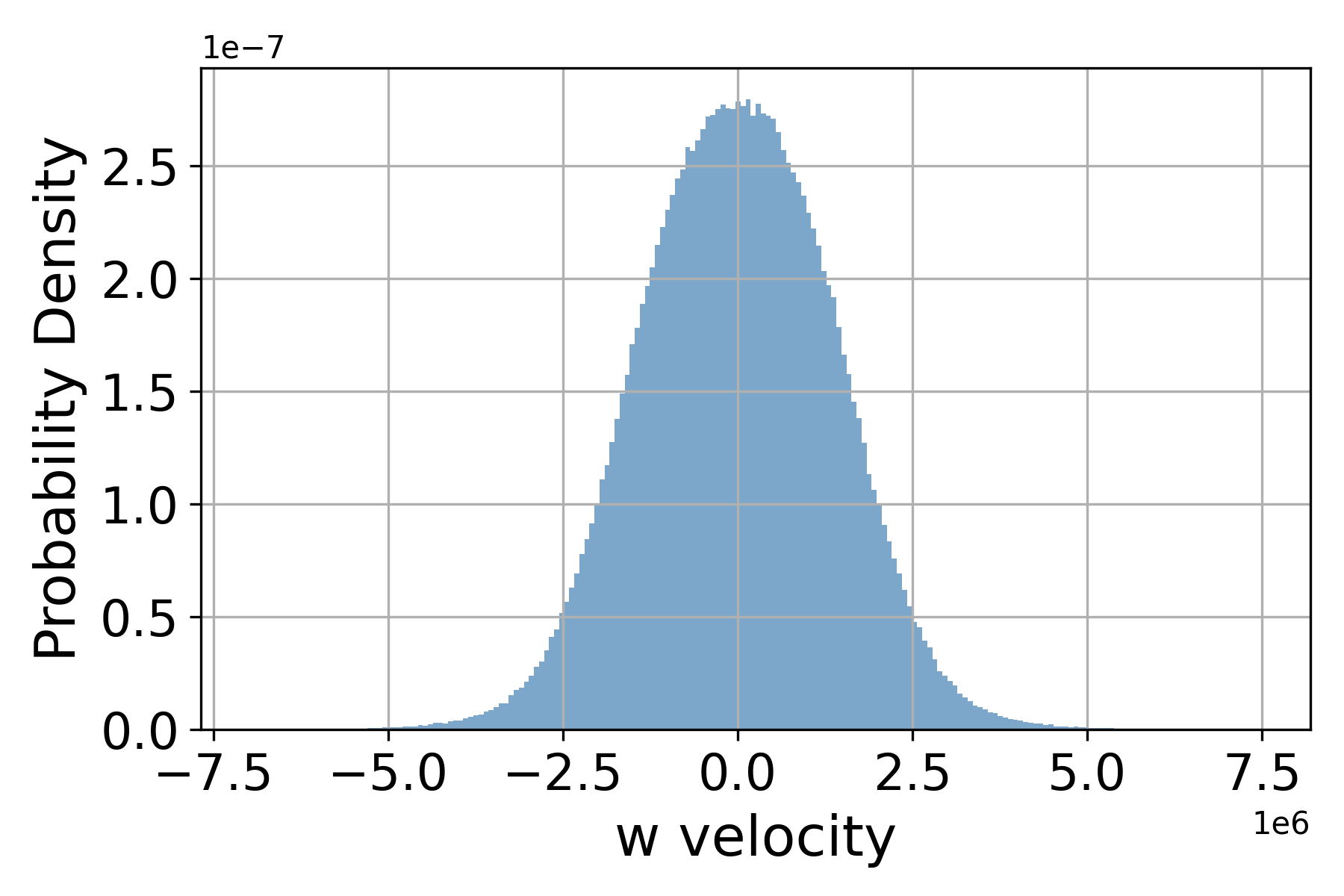}	
		\caption{
			Spatial and velocity distributions of inelastic collision events in case~1A up to 40~$\mu$s. 
			The left column shows the spatial locations $(x, y, z)$ where inelastic collisions occurred, 
			while the right column shows the corresponding post-collision electron velocity distributions $(u, v, w)$. 
			These results illustrate both the spatial localization of inelastic events and 
			the spread of electron velocities immediately after collisions.
		}
		\label{fig:inelastic_collision_data}
	\end{figure}
	
	\begin{figure}[htbp]
		\centering
		\includegraphics[width=0.85\linewidth]{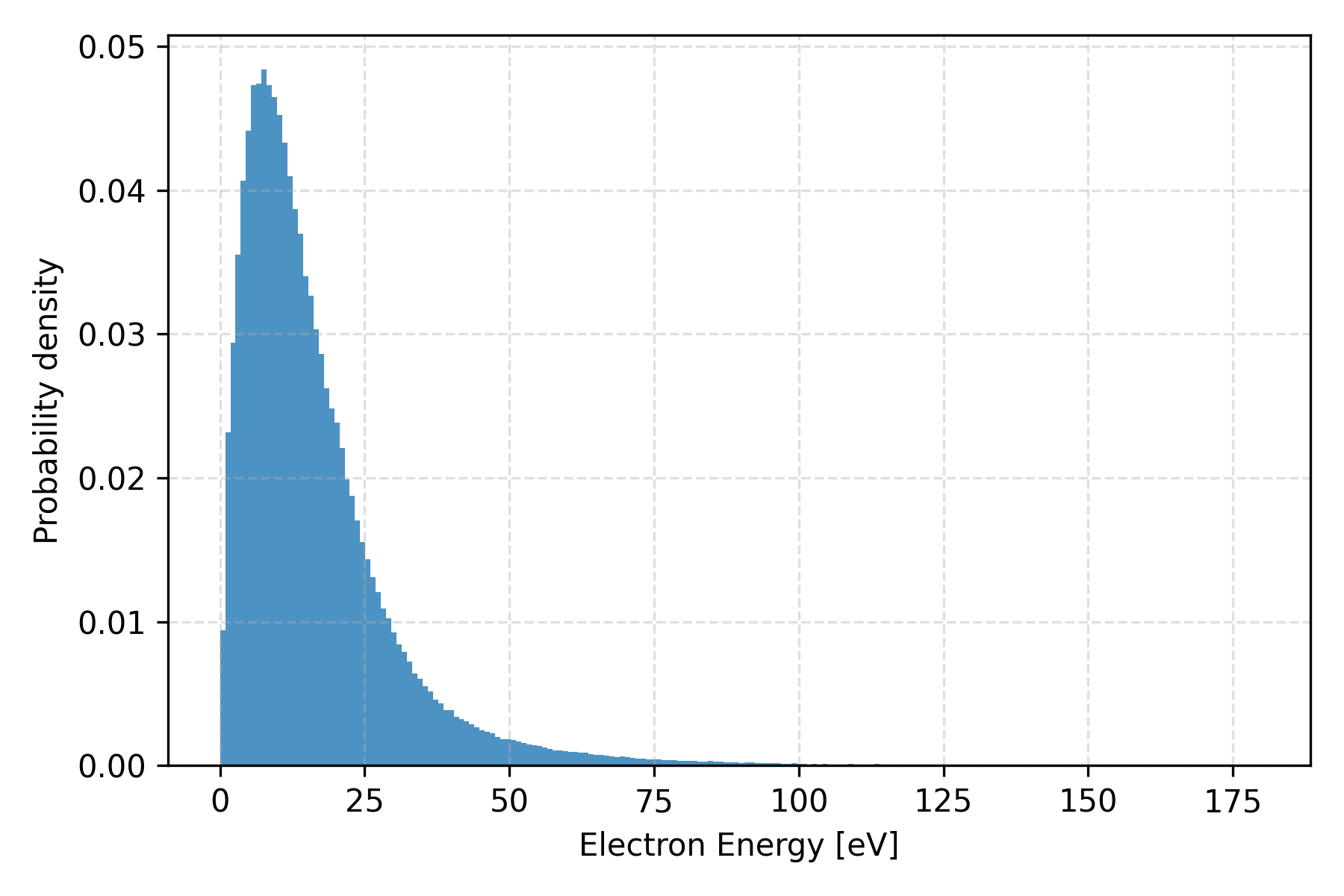}
		\caption{Energy distribution of electrons following inelastic collision events in case~1A up to 40~$\mu$s. 
			The figure presents the post-collision electron kinetic energy spectrum, 
			illustrating the characteristic spread in electron energies produced by inelastic processes. 
			The mean post-collision electron energy is 15.982~eV.}
		\label{fig:inelastic_collision_data-energy}
	\end{figure}
	
%

	\end{document}